\documentclass[12pt]{article}
\usepackage{fullpage}
\usepackage{psfig}
\usepackage{graphics}
\usepackage{amsmath,amssymb,bm}

\allowdisplaybreaks[1]
\def\be#1\ee{\begin{equation}#1\end{equation}}
\def\bal#1\eal{\begin{align}#1\end{align}}
\def\bat#1\eat{\begin{alignat}{2}#1\end{alignat}}
\def\bmu#1\emu{\begin{multline}#1\end{multline}}
\def\bga#1\ega{\begin{gather}#1\end{gather}}
\newcommand{\ba}{\begin{array}}
\newcommand{\ea}{\end{array}}
\newcommand{\n}{\notag}

\newcommand{\abs}[1]{\lvert#1\rvert}
\newcommand{\supfi}[1]{{}^{\,#1\!}}

\renewcommand{\d}{\partial}
\renewcommand{\bf}{\mathbf}
\renewcommand{\cal}{\mathcal}

\newcommand{\ssc}{\scriptscriptstyle}
\newcommand{\ds}{\displaystyle}

\title{\textbf{Fine and hyperfine structure \\
in different bound systems}}

\author{Axel Weber\thanks{Electronic address: 
\texttt{axel@ifm.umich.mx}} \\
\normalsize \em Instituto de F\'{\i}sica y Matem\'aticas, \\[-1mm] 
\normalsize \em Universidad Michoacana de San Nicol\'as de Hidalgo, \\[-1mm]
\normalsize \em Edificio C-3, Ciudad Universitaria, A. Postal 2-82, \\[-1mm]
\normalsize \em 58040 Morelia, Michoac\'an, Mexico}

\date{\normalsize August 31, 2005}

\begin{document}

\maketitle

\begin{abstract}
We demonstrate that the generalized Gell-Mann--Low theorem permits for
a systematic expansion around the nonrelativistic limit when applied
to bound states in the Wick-Cutkosky model, Yukawa theory, and QED
(in Coulomb gauge). We apply this expansion to obtain new results for the 
fine and hyperfine structure of bound states in the cases of the Wick-Cutkosky 
model and Yukawa theory, and reproduce correctly the fine and hyperfine 
structure of hydrogenic systems.
\end{abstract}

\section{Introduction}

In this paper we present a method to extract the fine and hyperfine structure,
presently to lowest order, of bound systems appearing in general relativistic 
quantum field theories in a systematic, consistent and transparent way.
It is based on the application of a generalization of the Gell-Mann--Low
theorem \cite{Web00} which generates an effective Hamiltonian called
the Bloch-Wilson Hamiltonian, for the dynamics of the two constituents.
The Bloch-Wilson Hamiltonian consists of the relativistic kinetic energy
of the constituents and an effective potential that can be obtained
explicitly in a perturbative expansion in powers of the coupling constant,
derived from the Dyson series for the adiabatic evolution operator.
At every perturbative order, the effective potential can be expanded in
powers of the relative momentum, which, together with the corresponding
expansion of the relativistic kinetic energy, leads to a systematic
expansion around the nonrelativistic limit.

We will demonstrate this procedure in three different theories, a scalar
model with cubic couplings (the ``Wick-Cutkosky model''), Yukawa theory,
and quantum electrodynamics (QED). The Bloch-Wilson Hamiltonian for two
of these theories has been carefully derived to lowest nontrivial order
(in the expansion in powers of the coupling constant) and used for 
numerical bound state calculations before \cite{WL02, WL05}.
The Bloch-Wilson Hamiltonian for QED in the Coulomb gauge has been
discussed very briefly in Ref.\ \cite{Web03}, and will be treated in much
more detail in a future publication. In each case, we will take the
exchanged boson to be massless in order to have analytic expressions
in the nonrelativistic limit and for the lowest-order fine and hyperfine
structure.

This procedure should be compared with the historical calculation of
the complete fine and hyperfine structure in hydrogenic systems. Ref.\
\cite{BS77} gives a good overview of these calculations which are based on
the Breit equation \cite{Bre29} and have not been completed until 1951. A 
more systematic treatment is provided by the Bethe-Salpeter equation 
\cite{BS51, GL51} in the perturbative expansion devised by Salpeter 
\cite{Sal52}. However, Salpeter's expansion begins at zero order with an 
instantaneous interaction of the constituents which is natural in QED, 
particularly in the Coulomb gauge, but not in the case of spinless boson 
exchange as in the Wick-Cutkosky model and Yukawa theory. In fact, the 
analogue in the Bethe-Salpeter approach to the procedure outlined above 
would be to consider the lowest-order approximation to the Bethe-Salpeter 
kernel, the so-called ladder approximation, and expand the equation in this
approximation around the nonrelativistic limit. However, even in the
simplest case of the Wick-Cutkosky model \cite{WC54} this procedure leads to 
a ``curious'' term of the unexpected order $\alpha^3 \ln \alpha$ \cite{FH73}.
The full fine structure in this model and in Yukawa theory have, to our
knowledge, never been calculated.

In the next two sections of this paper, we will present the Bloch-Wilson
Hamiltonians for the field theories mentioned above. In fact, we will
discuss two different types of effective Hamiltonians, in Section 2 the 
Bloch-Wilson Hamiltonian used before in Refs.\ \cite{WL02, WL05, Web03} 
which generalizes
in a natural way Bloch's formulation of degenerate perturbation theory 
\cite{Blo58}, and in Section 3 a hermitian version used before by Wilson 
\cite{Wil70} and, independently, by Gari \textit{et al.} \cite{GH76} 
and by Kr\"uger and Gl\"ockle \cite{KG99}, building on earlier work by 
Okubo \cite{Oku54}. As will become clear in Section 4, the two types of
effective Hamiltonians differ by an antihermitian term in the expansion
around the nonrelativistic limit which is present in the first type of
effective Hamiltonians (except for the case of QED) but obviously not
in the second. In Section 4, we will describe how to calculate the
fine and hyperfine structure to lowest order starting from these
effective Hamiltonians, relegating the more technical aspects of the
calculations to Appendix \ref{formul}. In particular, we will use second-order
time-independent perturbation theory to calculate the contributions
originating from the antihermitian term. An alternative method based on
a Foldy-Wouthuysen-type transformation from a nonhermitian to a 
hermitian Hamiltonian is detailed in Appendix \ref{FWtrans}. We will
also comment on the possible physical significance of the antihermitian
term. In Section 5, 
we discuss the results for the fine and hyperfine structure in the
different theories and compare in particular with the numerical solutions 
\cite{WL02, WL05} for the Bloch-Wilson Hamiltonian (to lowest nontrivial 
order in the expansion in powers of the coupling constant) without the 
expansion around the nonrelativistic limit, and with the well-known
fine and hyperfine structure in hydrogenic systems.

\section{The Bloch-Wilson Hamiltonian}

We will begin by stating the generalization of the Gell-Mann--Low theorem
proved in Ref.\ \cite{Web00}. To this end, some general notations have to be
introduced first: the full Hamiltonian $H$ of the field theory under
consideration is decomposed into a ``free'' Hamiltonian $H_0$ (typically
the one describing free particles) and an interaction Hamiltonian $H_1$,
$H = H_0 + H_1$. The adiabatic evolution operator $U_\epsilon$ maps a state 
$| \phi \rangle$ in the Fock space $\cal{F}$ in which $H_0$ acts, to the 
state $| \psi \rangle$ that results from evolving $| \phi \rangle$ with 
the Schr\"odinger equation corresponding to the adiabatic Hamiltonian
\be
H_\epsilon (t) = H_0 + e^{- \epsilon \abs{t}} H_1 \:, \qquad \epsilon > 0 \:,
\ee
from $t \to - \infty$ to $t = 0$. The adiabatic evolution operator has
the perturbative expansion
\be
U_\epsilon = \sum_{n=0}^\infty \frac{(-i)^n}{n!}
\int_{-\infty}^0 dt_1 \cdots \int_{-\infty}^0 dt_n \, e^{-\epsilon (\abs{t_1} 
+ \ldots + \abs{t_n})} \, T [ H_1 (t_1) \cdots H_1 (t_n) ] \:,
\label{dyson}
\ee
the well-known Dyson series, where
\be
H_1 (t) = e^{i H_0 t} H_1 \, e^{-i H_0 t} \:.
\ee

For the generalized Gell-Mann--Low theorem, a linear $H_0$-invariant subspace 
$\Omega_0 \subseteq \cal{F}$, $H_0 \Omega_0 \subseteq \Omega_0$, is fixed
and the corresponding orthogonal projection operator $P_0: \cal{F} \to 
\Omega_0$ introduced. The theorem asserts that, if the operator 
\be
U_B = \lim_{\epsilon \to 0} \, U_\epsilon (P_0 U_\epsilon P_0)^{-1} :
\Omega_0 \to U_B (\Omega_0) \equiv \Omega \:, \label{defUB}
\ee
the Bloch-Wilson operator, is well-defined in $\Omega_0$, then the image
subspace $\Omega$ is invariant under $H$, $H \Omega
\subseteq \Omega$. By its definition, $U_B$ is ``normalized'' to
\be
P_0 U_B = \bf{1}_{\Omega_0} = \left. P_0 \right|_{\Omega_0} \:, 
\label{normal}
\ee
consequently the inverse of $U_B$ is $\left. P_0 \right|_\Omega$.
This normalization naturally generalizes the one used in the original
Gell-Mann--Low theorem \cite{GL51} and corresponds to the one used by
Bloch in his formulation of degenerate perturbation theory \cite{Blo58}.
Note that we have changed the notation for the Bloch-Wilson operator
compared to our previous works to emphasize this fact, for reasons to
become clear later. The existence of the operator $U_B$ and the
assertion of the theorem are to be understood order by order in the
corresponding formal power series originating from Eq.\ \eqref{dyson},
as is the case for the original Gell-Mann--Low theorem.

The fact that $\Omega$ is $H$-invariant is equivalent to the
diagonalizability of $H$ in $\Omega$, so that part of the eigenvalue
problem of $H$ can be solved in $\Omega$. It is convenient to
similarity transform the eigenvalue problem back from $\Omega$ to
the usually more manageable subspace $\Omega_0$. Hence we define the
effective or Bloch-Wilson Hamiltonian
\be
H_B = P_0 H U_B : \Omega_0 \to \Omega_0 \:.
\ee
We can use Eq.\ \eqref{normal} and $P_0 H_0 = H_0 P_0$ as a consequence of
the $H_0$-invariance of $\Omega_0$, to bring $H_B$ into the form
\bal
H_B &= P_0 H_0 U_B + P_0 H_1 U_B \n \\
&= H_0 P_0 + P_0 H_1 U_B \:. \label{HBdec}
\eal
The eigenvalue problem of $H_B$ in $\Omega_0$ is equivalent to the one of
$H$ restricted to $\Omega$ with the identical eigenvalues. The corresponding
eigenstates are mapped into each other by $U_B$ and $P_0$.

For the following bound state calculations we will take $\Omega_0$ to be
the subspace of all states of the constituents as free particles (under
$H_0$), usually in a momentum eigenstate basis. In this case, the first
part $H_0 P_0$ in Eq.\ \eqref{HBdec} just represents the kinetic energy
of the constituents, consequently the second part defines an effective
potential for their interaction (although it also contains radiative
corrections to the vacuum energy and the masses of the constituents) given
in terms of a power series in the coupling constant by inserting the
Dyson series \eqref{dyson}.
The operator $P_0$ projects out from the true $H$-eigenstates in $\Omega$
the component with the particle numbers corresponding to the constituents,
hence we can think of the corresponding $H_B$-eigenstates as wave
functions of the constituents, the higher Fock space components (with higher 
particle numbers)
being generated by the application of $U_B$. For bound states, it is these
constituent wave functions that have to be normalizable, while the
corresponding $H$-eigenstates will usually not be normalizable with 
respect to the standard Fock space scalar product.

In the following, we will calculate $H_B$ to lowest nontrivial order for 
two-particle subspaces. In all field theories to be considered, the 
interaction Hamiltonian $H_1$ changes the particle number so that 
\be
P_0 H_1 (t) P_0 = 0 \:. \label{pnchange}
\ee
In this case, the lowest-order nontrivial contribution to $H_B$ arises
from the first-order term in the expansion \eqref{dyson}, and we can
write
\be
H_B = H_0 P_0 - i \int_{-\infty}^0 dt \, e^{- \epsilon \abs{t}} P_0
H_1 (0) H_1 (t) P_0 + \cal{O} (H_1^4) \:, \label{HB}
\ee
where the limit $\epsilon \to 0$ is understood.

\subsection{Wick-Cutkosky model}

We will now present the Bloch-Wilson Hamiltonian $H_B$ for three
different theories. We start with the ``Wick-Cutkosky model'', a field
theory with three scalar fields $\phi_A$, $\phi_B$, and $\varphi$, the
latter taken to be massless, with interaction Hamiltonian
\be
H_1 = g \int d^3 x \bm{:} \left[ \phi_A^\dagger (\bf{x}) 
\phi_A (\bf{x}) + \phi_B^\dagger (\bf{x}) 
\phi_B (\bf{x}) \right] \varphi (\bf{x}) \bm{:} \:. \label{model}
\ee
We consider states with one $A$ and one $B$ scalar. The Bloch-Wilson
Hamiltonian for the corresponding subspace was calculated in Ref.\
\cite{WL02}, with matrix elements
\bal
\langle \bf{p}_A, \bf{p}_B | H_B | \bf{p}_A', \bf{p}_B' \rangle
&= \left( \sqrt{m_A^2 + \bf{p}_A^2} + \sqrt{m_B^2 + \bf{p}_B^2} \right)
(2 \pi)^3 \delta (\bf{p}_A - \bf{p}_A') (2 \pi)^3 
\delta (\bf{p}_B - \bf{p}_B') \n \\
&\phantom{=} - \frac{g^2}{\sqrt{2 E_{\bf{p}_A}^A \, 2 E_{\bf{p}_B}^B \,
2 E_{\bf{p}_A'}^A \, 2 E_{\bf{p}_B'}^B}} \n \\
&\phantom{=} \times \frac{1}
{2 \abs{\bf{p}_A - \bf{p}_A'}} \left( \frac{1}{E_{\bf{p}_A}^A
+ \abs{\bf{p}_A - \bf{p}_A'} - E_{\bf{p}_A'}^A}
+ \frac{1}{E_{\bf{p}_B}^B + \abs{\bf{p}_B - \bf{p}_B'} - 
E_{\bf{p}_B'}^B} \right) \n \\[2mm]
&\phantom{=} \hspace{5cm} \times (2 \pi)^3 \delta (\bf{p}_A
+ \bf{p}_B - \bf{p}_A' - \bf{p}_B') \label{HBWC}
\eal
in a momentum eigenstate basis with the nonrelativistic normalization
\be
\langle \bf{p}_A, \bf{p}_B | \bf{p}_A', \bf{p}_B' \rangle
= (2 \pi)^3 \delta (\bf{p}_A - \bf{p}_A') (2 \pi)^3 
\delta (\bf{p}_B - \bf{p}_B') \:.
\ee
We have introduced the shorthands
\be
E_{\bf{p}}^{A,B} = \sqrt{m_{A,B}^2 + \bf{p}^2}
\ee
for the relativistic kinetic energies in Eq.\ \eqref{HBWC}. To arrive at
the form \eqref{HBWC}, the vacuum energy (including its lowest-order
radiative corrections) was subtracted, and radiative corrections to the
kinetic energy have been absorbed in a renormalization of the masses.
The renormalization procedure was analyzed in detail in Ref.\ \cite{WL05} 
from the present not manifestly covariant point of view, and the appearing 
one- and two-loop expressions have been shown to coincide with the ones of
usual covariant perturbation theory. The renormalized masses are denoted
here as $m_{A,B}$. Making use of the overall momentum conservation 
$\bf{p}_A + \bf{p}_B = \bf{p}_A' + \bf{p}_B'$, the dynamics can be reduced
to the center-of-mass system $\bf{p}_A + \bf{p}_B = \bf{0}$. One then has 
for the corresponding Schr\"odinger equation, with ${\bf p} = {\bf p}_A = 
- {\bf p}_B$,
\bmu
\left( \sqrt{m_A^2 + \bf{p}^2} + \sqrt{m_B^2 + \bf{p}^2} \right) 
\phi (\bf{p}) - \frac{g^2}{\sqrt{2 E_{\bf{p}}^A \,
2 E_{\bf{p}}^B}} \int \frac{d^3 p'}{(2 \pi)^3} \, 
\frac{1}{\sqrt{2 E_{\bf{p}'}^A \, 2 E_{\bf{p}'}^B}} \, \\
\times \frac{1}{2 \abs{\bf{p} - \bf{p}'}} \left( \frac{1}{E_{\bf{p}}^A
+ \abs{\bf{p} - \bf{p}'} - E_{\bf{p}'}^A}
+ \frac{1}{E_{\bf{p}}^B + \abs{\bf{p} - \bf{p}'} - 
E_{\bf{p}'}^B} \right) \phi (\bf{p}') = E \phi (\bf{p}) \:, 
\label{schrWC}
\emu
with the wave function $\phi (\bf{p})$ defined as
\be
\langle \bf{p}_A, \bf{p}_B | \phi \rangle = \phi (\bf{p}_A) (2 \pi)^3
\delta (\bf{p}_A + \bf{p}_B) \:.
\ee

\subsection{Yukawa theory}

As our second example, we consider Yukawa theory, consisting of two
Dirac fields $\psi_A$, $\psi_B$, a massless scalar field $\varphi$, and
the interaction Hamiltonian
\be
H_1 = g \int d^3 x \bm{:} \left[ \bar{\psi}_A (\bf{x}) \psi_A (\bf{x}) + 
\bar{\psi}_B (\bf{x}) \psi_B (\bf{x}) \right] \varphi (\bf{x}) \bm{:} \:.
\ee
The Bloch-Wilson Hamiltonian for bound states of one $A$ and one $B$
fermion was calculated in Ref.\ \cite{WL05}. The matrix elements are
\bal
\lefteqn{\langle \bf{p}_A, r; \bf{p}_B, s | H_B 
| \bf{p}'_A, r'; \bf{p}'_B, s' \rangle} \hspace{1cm} \n \\[2mm]
&= \left( \sqrt{m_A^2 + \bf{p}_A^2} + \sqrt{m_B^2 + \bf{p}_B^2}
\right) (2 \pi)^3 \delta (\bf{p}_A - \bf{p}'_A) \delta_{r r'} (2 \pi)^3
\delta (\bf{p}_B - \bf{p}'_B) \delta_{s s'} \n \\
&\phantom{=} - \frac{g^2}{\sqrt{2 E_{\bf{p}_A}^A \, 2 E_{\bf{p}_B}^B \, 
2 E_{\bf{p}'_A}^A \, 2 E_{\bf{p}'_B}^B}} \n \\
&\phantom{=} \times
\frac{1}{2 \abs{\bf{p}_A - \bf{p}'_A}} \left( \frac{1}{E_{\bf{p}_A}^A +
\abs{\bf{p}_A - \bf{p}'_A} - E_{\bf{p}'_A}^A} + \frac{1}{E_{\bf{p}_B}^B +
\abs{\bf{p}_B - \bf{p}'_B} - E_{\bf{p}'_B}^B} \right) \n \\[2mm]
&\phantom{=} \times \left[ \bar{u}_A (\bf{p}_A, r) \, u_A (\bf{p}'_A, r') 
\right] \left[ \bar{u}_B (\bf{p}_B, s) \, u_B (\bf{p}'_B, s') \right]
(2 \pi)^3 \delta (\bf{p}_A + \bf{p}_B - \bf{p}'_A - \bf{p}'_B) \:. 
\label{HBY}
\eal
The parameters $r, s$ with possible values $1, 2$ in the momentum
eigenstates $| \bf{p}_A, r; \bf{p}_B, s \rangle$ describe the spin
orientations of fermions $A$ and $B$, respectively. The Dirac spinors 
$u_A (\bf{p}, r)$ are related to Pauli spinors $\chi_r$ via
\be
u_A (\bf{p}, r) = \sqrt{E^A_{\bf{p}} + m_A} 
\left( \ba{c} \chi_r \\[4mm]
\ds \frac{ \bf{p} \cdot \bm{\sigma} }{E^A_{\bf{p}} + m_A} \, 
\chi_r \ea \right) \:,
\ee
with the Pauli spinors normalized to
\be
\chi_r^\dagger \chi_s = \delta_{rs} \:.
\ee
Note that Eqs.\ \eqref{HBWC} and \eqref{HBY} only differ by the Dirac
spinor products (and the additional spin indices).

The remarks after
Eq.\ \eqref{HBWC} apply equally to Eq.\ \eqref{HBY}. In the center-of-mass
system, we can write the corresponding Schr\"odinger equation
conveniently in Pauli spinor form as
\bal
\lefteqn{\left( \sqrt{m_A^2 + \bf{p}^2} + \sqrt{m_B^2 + \bf{p}^2} \right)
\phi (\bf{p})} \hspace{1cm} \n \\
&- g^2 \sqrt{\frac{E^A_{\bf{p}} + m_A}{2 E^A_{\bf{p}}} \,
\frac{E^B_{\bf{p}} + m_B}{2 E^B_{\bf{p}}}} \int \frac{d^3 p'}{(2 \pi)^3} 
\sqrt{\frac{E^A_{\bf{p}'} + m_A}{2 E^A_{\bf{p}'}} \,
\frac{E^B_{\bf{p}'} + m_B}{2 E^B_{\bf{p}'}}} \n \\
&\times \frac{1}{2 \abs{\bf{p} - \bf{p}'}} \left( \frac{1}{E_{\bf{p}}^A +
\abs{\bf{p} - \bf{p}'} - E_{\bf{p}'}^A} + \frac{1}{E_{\bf{p}}^B +
\abs{\bf{p} - \bf{p}'} - E_{\bf{p}'}^B} \right) \n \\[2mm]
&\times \left(
1 - \frac{\bf{p} \cdot \bm{\sigma}_A}{E^A_{\bf{p}} + m_A} \,
\frac{\bf{p}' \cdot \bm{\sigma}_A}{E^A_{\bf{p}'} + m_A} \right)
\left( 1 - \frac{\bf{p} \cdot \bm{\sigma}_B}{E^B_{\bf{p}} + m_B} \,
\frac{\bf{p}' \cdot \bm{\sigma}_B}{E^B_{\bf{p}'} + m_B} \right)
\phi (\bf{p}') = E \phi (\bf{p}) \:, \label{schrY}
\eal
where the spinorial wave function $\phi ({\bf p})$ in the center-of-mass
system is defined as
\be
\phi (\bf{p}_A) (2 \pi)^3 \delta (\bf{p}_A + \bf{p}_B) 
= \sum_{r,s} \langle \bf{p}_A, r; \bf{p}_B, s | \phi \rangle 
\left[ \chi_r \otimes \chi_s \right] \:, \label{wavefdef}
\ee
and $\bm{\sigma}_A$ ($\bm{\sigma}_B$) is understood to act on $\chi_r$ 
($\chi_s$) only.

\subsection{Quantum electrodynamics}

As our final example, we turn to quantum electrodynamics (QED). We
consider the Coulomb gauge for the massless gauge field $A$ which we
couple to two different Dirac fields $\psi_A$ and $\psi_B$ with 
opposite electric charges $\pm e$. Then the interaction Hamiltonian is
\bmu
H_1 = e \int d^3 x \bm{:} \left[ \bar{\psi}_A (\bf{x}) \gamma^\mu
\psi_A (\bf{x}) - \bar{\psi}_B (\bf{x}) \gamma^\mu \psi_B (\bf{x}) \right] 
A_\mu^{\text{tr}} (\bf{x}) \bm{:} \\
+ \frac{e^2}{8 \pi} \int d^3 x \, d^3 y \bm{:} \frac{ \left[ 
\bar{\psi}_A (\bf{x}) \gamma^0 \psi_A (\bf{x}) - \bar{\psi}_B (\bf{x}) 
\gamma^0 \psi_B (\bf{x}) \right] \left[ \bar{\psi}_A (\bf{y}) \gamma^0 
\psi_A (\bf{y}) - \bar{\psi}_B (\bf{y}) \gamma^0 \psi_B (\bf{y}) \right]}
{\abs{\bf{x} - \bf{y}}} \bm{:} \:,
\emu
where $A^{\text{tr}}$ denotes the spatially transverse part of the gauge
field, the dyamical degrees of freedom remaining from $A$ after
the Coulomb gauge fixing. They are most simply characterized by
\be
A^{\text{tr}}_0 (\bf{k}) = 0 \:, \qquad \bf{k} \cdot \bf{A}^{\text{tr}}
(\bf{k}) = 0
\ee
for the Fourier coefficients $A^{\text{tr}}_\mu (\bf{k})$ of
$A^{\text{tr}}_\mu (\bf{x})$, and can also be projected out of $A$ by
applying (the negative of) the transverse Kronecker delta 
$\delta^{\text{tr}} (\bf{k})$,
\be
\delta^{\text{tr}}_{0 0} (\bf{k}) = \delta^{\text{tr}}_{0 i} (\bf{k}) =
\delta^{\text{tr}}_{i 0} (\bf{k}) = 0 \:, \qquad
\delta^{\text{tr}}_{i j} (\bf{k}) = \delta_{i j} - \frac{k_i k_j}{\bf{k}^2}
\:. \label{defdeltr}
\ee

The Bloch-Wilson Hamiltonian for states with one $A$ and one $B$ fermion
can be calculated in strict analogy to the calculations in Refs.\
\cite{WL02, WL05} for the Wick-Cutkosky model and Yukawa theory. Details
will be given in a future publication. The result for the matrix elements
of $H_B$ is
\bal
\lefteqn{\langle \bf{p}_A, r; \bf{p}_B, s | H_B 
| \bf{p}'_A, r'; \bf{p}'_B, s' \rangle} \hspace{1cm} \n \\[2mm]
&= \left( \sqrt{m_A^2 + \bf{p}_A^2} + \sqrt{m_B^2 + \bf{p}_B^2}
\right) (2 \pi)^3 \delta (\bf{p}_A - \bf{p}'_A) \delta_{r r'} (2 \pi)^3
\delta (\bf{p}_B - \bf{p}'_B) \delta_{s s'} \n \\
&\phantom{=} - \frac{e^2}{\sqrt{2 E_{\bf{p}_A}^A \, 2 E_{\bf{p}_B}^B
\, 2 E_{\bf{p}_A'}^A \, 2 E_{\bf{p}_B'}^B}} \n \\
&\phantom{=} \times \Bigg[ \frac{1}{\left( \bf{p}_A - \bf{p}_A' \right)^2}
\left[ \bar{u}_A (\bf{p}_A, r) \gamma^0 u_A (\bf{p}_A', r') \right] 
\left[ \bar{u}_B (\bf{p}_B, s) \gamma^0 u_B (\bf{p}_B', s') \right] \n \\
&\phantom{=} - \frac{1}{2 \abs{\bf{p}_A - \bf{p}_A'}} \left( \frac{1}{
E_{\bf{p}_A}^A + \abs{\bf{p}_A - \bf{p}_A'} - E_{\bf{p}_A'}^A} + \frac{1}{
E_{\bf{p}_B}^B + \abs{\bf{p}_B - \bf{p}_B'} - E_{\bf{p}_B'}^B} \right) \n \\
&\phantom{=} \hspace{5mm} \times \left[ \bar{u}_A (\bf{p}_A, r) 
\gamma^\mu u_A ({\bf p}_A', r') \right] 
\left( \sum_{\lambda = 1}^2 \varepsilon_\mu^{(\lambda)} 
(\bf{p}_A - \bf{p}_A') \, \varepsilon_\nu^{(\lambda) \ast} 
(\bf{p}_A - \bf{p}_A') \right) \n \\
&\phantom{=} \hspace{5mm} \times \left[ \bar{u}_B (\bf{p}_B, s) 
\gamma^\nu u_B (\bf{p}_B', s') \right] \Bigg] (2 \pi)^3 
\delta (\bf{p}_A + \bf{p}_B - \bf{p}'_A - \bf{p}'_B) \:, \label{HBQED}
\eal
where $\varepsilon^{(\lambda)} (\bf{k})$ for $\lambda = 1,2$ are the 
spatially transverse polarization vectors, so that
\be
\sum_{\lambda = 1}^2 \varepsilon_\mu^{(\lambda)} (\bf{k}) \, 
\varepsilon_\nu^{(\lambda) \ast} (\bf{k}) 
= \delta^{\text{tr}}_{\mu \nu} (\bf{k}) \:.
\ee

The Schr\"odinger equation corresponding to the Bloch-Wilson Hamiltonian
\eqref{HBQED} in the center-of-mass system is written in Pauli spinor
form as
\bal
\lefteqn{\left( \sqrt{m_A^2 + \bf{p}^2} + \sqrt{m_B^2 + \bf{p}^2} \right)
\phi (\bf{p})} \n \\
&- e^2 \sqrt{
\frac{E^A_{\bf{p}} + m_A}{2 E^A_{\bf{p}}} \,
\frac{E^B_{\bf{p}} + m_B}{2 E^B_{\bf{p}}}} \int \frac{d^3 p'}{(2 \pi)^3} 
\sqrt{\frac{E^A_{\bf{p}'} + m_A}{2 E^A_{\bf{p}'}} \,
\frac{E^B_{\bf{p}'} + m_B}{2 E^B_{\bf{p}'}}} \n \\
&\times \Bigg[ \frac{1}{\left( \bf{p} - \bf{p}' \right)^2}
\left( 1 + \frac{\bf{p} \cdot \bm{\sigma}_A}{E^A_{\bf{p}} + m_A} \,
\frac{\bf{p}' \cdot \bm{\sigma}_A}{E^A_{\bf{p}'} + m_A} \right)
\left( 1 + \frac{\bf{p} \cdot \bm{\sigma}_B}{E^B_{\bf{p}} + m_B} \,
\frac{\bf{p}' \cdot \bm{\sigma}_B}{E^B_{\bf{p}'} + m_B} \right) \n \\
&+ \frac{1}{2 \abs{\bf{p} - \bf{p}'}} \left( \frac{1}{E_{\bf{p}}^A +
\abs{\bf{p} - \bf{p}'} - E_{\bf{p}'}^A} + \frac{1}{E_{\bf{p}}^B +
\abs{\bf{p} - \bf{p}'} - E_{\bf{p}'}^B} \right) \n \\
&\times \left( 
\frac{(\bf{p} \cdot \bm{\sigma}_A) \sigma_A^i}{E^A_{\bf{p}} + m_A}
+ \frac{\sigma_A^i (\bf{p}' \cdot \bm{\sigma}_A)}{E^A_{\bf{p}'} + m_A} \right)
\delta^{\text{tr}}_{i j} (\bf{p} - \bf{p}')
\left( \frac{(\bf{p} \cdot \bm{\sigma}_B) \sigma_B^j}{E^B_{\bf{p}} + m_B}
+ \frac{\sigma_B^j (\bf{p}' \cdot \bm{\sigma}_B)}{E^B_{\bf{p}'} + m_B} \right) 
\Bigg] \phi (\bf{p}') = E \phi (\bf{p}) \:, \label{schrQED}
\eal
with the spinorial wave function $\phi (\bf{p})$ defined as in Eq.\
\eqref{wavefdef}.

\section{The Okubo Hamiltonian \label{OHder}}

After these concrete examples, let us come back to the general formulae
for a moment. Our choice of $U_B$ and $H_B$ has the advantages of
relative simplicity and the
wave function interpretation of the $H_B$-eigenstates. However, since
$U_B$ is not unitary (this is maybe clearest for its inverse, 
$\left. P_0 \right|_\Omega$), the Bloch-Wilson Hamiltonian $H_B$ is in
general not hermitian, as we will see explicitly in the next section for our
specific examples. The nonhermiticity of the effective Hamiltonian 
might be a serious drawback in practical applications, although it must
be mentioned that it has not led to any problems in the numerical
calculations of Refs.\ \cite{WL02, WL05}. In any case, a simple possibility
is to replace $U_B$ by its unitary part
\be
U_W = U_B (U_B^\dagger U_B)^{-1/2} : \Omega_0 \to \Omega \label{defUW}
\ee
which maps $\Omega_0$ to the same subspace $\Omega$ as does $U_B$. Indeed,
\be
U_W^\dagger = (U_B^\dagger U_B)^{-1/2} \, U_B^\dagger 
= (U_B^\dagger U_B)^{1/2} \, U_B^{-1} = U_W^{-1} \:. \label{UWunit}
\ee
The corresponding effective Hamiltonian
\be
H_W = U_W^{-1} H U_W = (U_B^\dagger U_B)^{1/2} H_B (U_B^\dagger U_B)^{-1/2}
\label{defHW}
\ee
is then hermitian, as wished. We will refer to $U_W$ and $H_W$ in the 
following as the Okubo map and the Okubo Hamiltonian to distinguish them from 
$U_B$ and $H_B$. They were first introduced by Okubo \cite{Oku54} in a way 
that does not refer to the generalized Gell-Mann--Low theorem or the
adiabatic evolution operator. As a consequence, there is no 
$i \epsilon$-prescription for the energy denominators in Okubo's original 
formulation, and the relation to Feynman diagrams is not obvious.

The Okubo map was first applied to field theory by Wilson in Ref.\
\cite{Wil70}, who however did not build upon Okubo's work. Later, 
Gari \textit{et al.} and Kr\"uger and Gl\"ockle used the Okubo map 
introduced in Ref.\ \cite{Oku54}, however without any reference 
to Wilson's earlier work.
To complete this historical interlude, it was apparently realized by
Wilson's collaborators that the Okubo map generalizes the formulae
developed by Bloch for degenerate perturbation theory \cite{Blo58}.
Independently, the present author generalized Bloch's formulation to the
Bloch-Wilson map and subsequently realized its connection with a
generalization of the Gell-Mann--Low theorem \cite{Web00}, before
becoming aware of the older literature.

We will now derive the explicit second-order expression for $H_W$
analogous to Eq.\ \eqref{HB} for $H_B$, for the case of particle number
changing interactions so that Eq.\ \eqref{pnchange} holds. Equation
\eqref{defUB} implies, due to the unitarity of $U_\epsilon$, that
\be
U_B^\dagger U_B = (P_0 U_\epsilon^\dagger P_0)^{-1} 
(P_0 U_\epsilon P_0)^{-1} \:.
\ee
The expansion \eqref{dyson} then leads to
\bal
U_B^\dagger U_B &= P_0 + \int_{-\infty}^0 d t_1 \int_{-\infty}^{t_1} d t_2 
\, e^{-\epsilon (\abs{t_1} +\abs{t_2})} P_0 H_1 (t_1) H_1 (t_2) P_0 \n \\
&\phantom{=} \hspace{1cm} + \int_{-\infty}^0 d t_1 \int_{-\infty}^{t_1} d t_2 
\, e^{-\epsilon (\abs{t_1} +\abs{t_2})} P_0 H_1 (t_2) H_1 (t_1) P_0 \n \\
&= P_0 + \int_{-\infty}^0 d t_1 \int_{-\infty}^0 d t_2 \,
e^{-\epsilon (\abs{t_1} +\abs{t_2})} P_0 H_1 (t_1) H_1 (t_2) P_0 \:,
\label{UB2}
\eal
to second order in $H_1$. Now, from Eq.\ \eqref{defHW} we have for $H_W$, 
again to second order,
\bal
H_W &= H_0 P_0 - i \int_{-\infty}^0 d t \,
e^{-\epsilon \abs{t}} P_0 H_1 (0) H_1 (t) P_0 \n \\
&\phantom{=} \hspace{1cm} 
+ \frac{1}{2} \int_{-\infty}^0 d t_1 \int_{-\infty}^0 d t_2 \,
e^{-\epsilon (\abs{t_1} + \abs{t_2})} P_0 
\left[ H_1 (t_1) H_1 (t_2), H_0 \right] P_0 \:, \label{HWcom}
\eal
where Eqs.\ \eqref{HB} and \eqref{UB2} have been used. The commutator in this 
expression can be rewritten as
\bal
\left[ H_1 (t_1) H_1 (t_2), H_0 \right] &= 
\left[ H_1 (t_1), H_0 \right] H_1 (t_2)
+ H_1 (t_1) \left[ H_1 (t_2), H_0 \right] \n \\
&= i \frac{d}{d t_1} H_1 (t_1 ) H_1 (t_2 ) + i H_1 (t_1) \frac{d}{d t_2}
H_1 (t_2) \:, \label{commeval}
\eal
and integration by parts and the limit $\epsilon \to 0$ finally lead to
\bal
H_W &= H_0 P_0 - \frac{i}{2} \int_{-\infty}^0 d t \,
e^{-\epsilon \abs{t}} P_0 H_1 (0) H_1 (t) P_0 + \frac{i}{2} \int_{-\infty}^0 
d t \, e^{-\epsilon \abs{t}} P_0 H_1 (t) H_1 (0) P_0 \n \\
&= \frac{1}{2} ( H_B + H_B^\dagger ) \:, \label{HWherm}
\eal
i.e., $H_W$ is precisely the hermitian part of $H_B$ (to this order).

Concerning these calculations, let us remark that there are two formally
different representations for the inverse of $U_W$. From Eqs.\ \eqref{dyson},
\eqref{defUW} and \eqref{UB2}, we have
\bal
U_W &= P_0 -i \int_{-\infty}^0 d t \, e^{-\epsilon \abs{t}} H_1 (t) P_0
- \int_{-\infty}^0 d t_1 \int_{-\infty}^{t_1} d t_2 \, 
e^{-\epsilon (\abs{t_1} + \abs{t_2})} H_1 (t_1) H_1 (t_2) P_0 \n \\
&\phantom{=} \hspace{1cm} + \frac{1}{2} \int_{-\infty}^0 d t_1 
\int_{-\infty}^{t_1} d t_2 \, e^{-\epsilon (\abs{t_1} + \abs{t_2})} 
P_0 \left[ H_1 (t_1), H_1 (t_2) \right] P_0 \:,
\eal
to second order. Then the second-order expression for $U_W^\dagger$ is
different from the formal second-order expression for the inverse, the
latter turning out to be
\be
U_W^{-1} = P_0 + \frac{1}{2} \int_{-\infty}^0 d t_1 \int_{-\infty}^0 d t_2 
\, e^{-\epsilon (\abs{t_1} +\abs{t_2})} P_0 H_1 (t_1) H_1 (t_2) P_0 \:.
\label{UWinv}
\ee
However, it is straightforward to verify, by working consequently to second 
order, that both expressions lead to the same result when applied to an 
element of $\Omega = U_B \Omega_0$ , as they must according to the unitarity 
of $U_W$, Eq.\ \eqref{UWunit}. As a result of the $H$-invariance of $\Omega$, 
the corresponding expressions for $H_W$ are also formally identical to second 
order,
\be
H_W = U_W^\dagger H U_W = U_W^{-1} H U_W \:, \label{HWident}
\ee
hence the present formulation is consistent. The calculation of $H_W$
above implicitly uses the expression \eqref{UWinv}. Note, however, that 
identities like Eq.\ \eqref{commeval} are required to establish Eq.\ 
\eqref{HWident} (they are also required to establish the $H$-invariance of 
$\Omega$, i.e., the generalized Gell-Mann--Low theorem), and that an equality 
like Eq.\ \eqref{HWident} does \emph{not} hold for the $H_0$- or $H_1$-part 
of $H_W$ individually.

Equation \eqref{HWherm} allows to calculate the matrix elements of $H_W$ and 
the corresponding effective Schr\"odinger equation very easily given the
corresponding results for $H_B$. In particular, for all theories considered
here, the matrix elements of $H_B$ for the zero- and one-particle states
are diagonal and real, hence the results for the vacuum energy and the
renormalized masses including the lowest-order radiative corrections,
are identical for $H_B$ and $H_W$. For the Wick-Cutkosky model, the only
change in going from $U_B$ to $U_W$ in Eqs.\ \eqref{HBWC} and \eqref{schrWC},
is the replacement
\bmu
\frac{1}{2 \abs{\bf{p}_A - \bf{p}_A'}} \left( \frac{1}{E_{\bf{p}_A}^A
+ \abs{\bf{p}_A - \bf{p}_A'} - E_{\bf{p}_A'}^A}
+ \frac{1}{E_{\bf{p}_B}^B + \abs{\bf{p}_B - \bf{p}_B'} - 
E_{\bf{p}_B'}^B} \right) \\
\longrightarrow \frac{1}{2} \left( 
\frac{1}{\big( \bf{p}_A - \bf{p}_A' \big)^2 -
\big( E_{\bf{p}_A}^A - E_{\bf{p}_A'}^A \big)^2} 
+ \frac{1}{\big( \bf{p}_B - \bf{p}_B' \big)^2 -
\big( E_{\bf{p}_B}^B - E_{\bf{p}_B'}^B \big)^2} \right) \label{replace}
\emu
of the effective potential with its ``symmetric part''. In the case of
Yukawa theory, one has to take into account, in addition, that for the spinor 
products
\be
[ \bar{u}_A (\bf{p}_A', r') \, u_A (\bf{p}_A, r) ]^\ast
= \bar{u}_A (\bf{p}_A, r) \, u_A (\bf{p}_A', r')
\ee
holds. As a consequence, Eqs.\ \eqref{HBY} and \eqref{schrY} are unchanged 
for $H_W$, except for the replacement \eqref{replace}. The same is true for
Eqs.\ \eqref{HBQED} and \eqref{schrQED} in Coulomb gauge QED, due to an
analogous property of the spinor products and the corresponding property of 
$\delta^{\text{tr}}$ [see Eq.\ \eqref{defdeltr}].

\section{Expansion around the nonrelativistic limit}

It was emphasized in Refs.\ \cite{WL02, WL05} that it is very simple to 
obtain the nonrelativistic limit from the Bloch-Wilson Hamiltonian, by
just considering the leading terms for small relative momentum. In all
the theories considered in the previous section, one obtains in this way
a nonrelativistic Schr\"odinger equation with a Coulomb potential,
as a consequence of the massless exchanged boson. In the present
section, we will show that the next-to-leading terms in a systematic
expansion in powers of the relative momentum can be used to obtain the
(lowest-order) fine and hyperfine structure of bound states by application 
of time-independent Rayleigh-Schr\"odinger perturbation theory.

The lowest-order fine and hyperfine structure is of course well-known
for the case of QED. It was first obtained for hydrogenic systems 
\cite{BS77} from the Breit equation \cite{Bre29}. Note that the 
effective Schr\"odinger equation \eqref{schrQED} presented in the previous
section does not properly apply to positronium because there is an
additional contribution (the virtual annihilation graph) to the Bloch-Wilson
Hamiltonian in the case of a fermion-antifermion bound state. Nor does it
apply to hydrogen because of the anomalous $g$-factor of the proton,
while $g=2$ is implicit in Eq.\ \eqref{schrQED}. It is, however, appropriate
for the description of muonium, an antimuon-electron bound state. For the
other two cases, the Wick-Cutkosky model and Yukawa theory, the fine and
hyperfine structure have, to our knowledge, not been determined before.

\subsection{Momentum expansion \label{momexp}}

The crucial new feature arising in the fine structure for
the Wick-Cutkosky model and Yukawa theory is the contribution from the
retardation in the effective potential. Explicitly, the lowest orders
of a systematic expansion in powers of the relative momentum are, for $H_B$,
\bal
\frac{1}{|\bf{p} -\bf{p}'|} \, 
\frac{1}{E_{\bf{p}}^A + |\bf{p} - \bf{p}'| - E_{\bf{p}'}^A}
&= \frac{1}{\left( \bf{p} -\bf{p}' \right)^2} 
\bigg[ 1 - \frac{p + p'}{2 m_A} \frac{p - p'}{|\bf{p} - \bf{p}'|} \n \\
&\phantom{=} \hspace{2cm} + \left( \frac{p + p'}{2 m_A} 
\frac{p - p'}{|\bf{p} - \bf{p}'|}
\right)^2 + {\cal O} \left[ (p/m_A)^3 \right] \bigg] \:, \label{retard}
\eal
where $p \equiv |\bf{p}|$. The leading term corresponds to the Coulomb
potential, while the first-order, \emph{antihermitian} correction term
and the second-order hermitian term will turn out to contribute to the same 
order of the fine structure. On the other hand, for $H_W$ only the
hermitian part contributes, explicitly
\be
\frac{1}{\big( \bf{p} - \bf{p}' \big)^2 -
\big( E_{\bf{p}}^A - E_{\bf{p}'}^A \big)^2}
= \frac{1}{\left( \bf{p} -\bf{p}' \right)^2} 
\bigg[ 1 + \left( \frac{p + p'}{2 m_A} \frac{p - p'}{|\bf{p} - \bf{p}'|}
\right)^2 + {\cal O} \left[ (p/m_A)^4 \right] \bigg] \:. \label{retardh}
\ee
Since the antihermitian term in Eq.\ \eqref{retard} does contribute to the
lowest-order fine structure, the effective Hamiltonians $H_B$ and $H_W$
lead to \emph{different} results to this order, for the Wick-Cutkosky model
and Yukawa theory. This fact and its possible implications will be discussed 
further towards the end of this section and in the next section.

We will now present the complete results for the expansion in powers of
relative momentum to next-to-leading orders, for the different effective
Hamiltonians discussed in the previous section. For the Wick-Cutkosky
model and the Hamiltonian $H_B$, the expansion of the effective Schr\"odinger 
equation \eqref{schrWC} to next-to-leading order gives
\addtocounter{equation}{1}
\bal
& \left[ \frac{p^2}{2 m_r} - \left( \frac{p^4}{8 m_A^3} + \frac{p^4}{8 m_B^3}
\right) \right] \phi(\bf{p}) \tag{\theequation$a$} \label{WCkin} \\
&- \frac{g^2}{2 m_A \, 2 m_B} \int \frac{d^3 p'}{(2 \pi)^3} \, 
\frac{1}{\left( \bf{p} - \bf{p}' \right)^2}
\bigg[ 1 - \left( \frac{p^2 + p'^2}{4 m_A^2} +
\frac{p^2 + p'^2}{4 m_B^2} \right) \tag{\theequation$b$} \label{WCnloc} \\
&\phantom{-} \hspace{1.8cm} - \frac{p^2 - p'^2}{4 m_r |\bf{p} - \bf{p}'|} 
\tag{\theequation$c$} \label{WCreta} \\
&\phantom{-} \hspace{1.8cm} + \left( \frac{(p^2 - p'^2)^2}
{8 m_A^2 \left( \bf{p} - \bf{p}' \right)^2} + \frac{(p^2 - p'^2)^2}
{8 m_B^2 \left( \bf{p} - \bf{p}' \right)^2} \right) \bigg] 
\phi (\bf{p}') = (E - m_A - m_B) \phi (\bf{p}) \:, \tag{\theequation$d$} 
\label{WCreth}
\eal
where we have introduced the reduced mass
\be
m_r = \frac{m_A \, m_B}{m_A + m_B} \:.
\ee
Analogously, for the Yukawa theory and the Hamiltonian $H_B$, Eq.\
\eqref{schrY} leads to
\addtocounter{equation}{1}
\bal
& \left[ \frac{p^2}{2 m_r} - \left( \frac{p^4}{8 m_A^3} + \frac{p^4}{8 m_B^3}
\right) \right] \phi(\bf{p}) \tag{\theequation$a$} \label{Ykin} \\
&- g^2 \int \frac{d^3 p'}{(2 \pi)^3} \, 
\frac{1}{\left( \bf{p} - \bf{p}' \right)^2}
\bigg[ 1 - \left( \frac{p^2 + p'^2}{8 m_A^2} +
\frac{p^2 + p'^2}{8 m_B^2} \right) \tag{\theequation$b$} \label{Ynloc} \\
&\phantom{-} \hspace{5mm} - \frac{p^2 - p'^2}{4 m_r |\bf{p} - \bf{p}'|} 
\tag{\theequation$c$} \label{Yreta} \\
&\phantom{-} \hspace{5mm} + \left( \frac{(p^2 - p'^2)^2}
{8 m_A^2 \left( \bf{p} - \bf{p}' \right)^2}
+ \frac{(p^2 - p'^2)^2}{8 m_B^2 \left( \bf{p} - \bf{p}' \right)^2} \right) 
\tag{\theequation$d$} \label{Yreth} \\
&\phantom{-} \hspace{5mm} - \left( \frac{(\bf{p} \cdot \bm{\sigma}_A)
(\bf{p}' \cdot \bm{\sigma}_A)}{4 m_A^2} +
\frac{(\bf{p} \cdot \bm{\sigma}_B)
(\bf{p}' \cdot \bm{\sigma}_B)}{4 m_B^2} \right) \bigg] 
\phi (\bf{p}') = (E - m_A - m_B) \phi (\bf{p}) \:, \tag{\theequation$e$} 
\label{Yspin}
\eal
to next-to-leading order. Note the differences between the two effective
Schr\"odinger equations, a relative factor of 1/2 between Eqs.\
\eqref{WCnloc} and \eqref{Ynloc}, and the appearance in Eq.\ \eqref{Yspin} 
of the Pauli matrices acting on the Pauli spinors of $\phi (\bf{p}')$
[see Eq.\ \eqref{wavefdef}], both originating from the product of Dirac
spinors in Eq.\ \eqref{HBY}.

Turning now to QED in the Coulomb gauge, we can see directly from the
effective Schr\"odinger equation \eqref{schrQED} why there is no
contribution from the retardation to the lowest-order fine structure:
the instantaneous Coulomb interaction of the charge densities has no
retardation, and the interaction with retardation transmitted by the
spatially transverse photons carries additional powers of momentum, so
that retardation only contributes to higher orders. As a consequence, there 
is no antihermitian term in the expansion to next-to-leading order, and $H_B$
and $H_W$ lead to the same fine and hyperfine structure. 

In order to obtain a compact explicit expression for the next-to-leading
terms in the expansion, some algebra involving the Pauli matrices is
required. In particular, the identity
\be
(\bf{p} \cdot \bm{\sigma}_A) \bm{\sigma}_A + \bm{\sigma}_A
(\bf{p}' \cdot \bm{\sigma}_A) = \bf{p} + \bf{p}' + i \, \bm{\sigma}_A
\times (\bf{p} - \bf{p}')
\ee
is helpful in intermediate steps. The final result can be written as
\addtocounter{equation}{1}
\bal
& \left[ \frac{p^2}{2 m_r} - \left( \frac{p^4}{8 m_A^3} + \frac{p^4}{8 m_B^3}
\right) \right] \phi(\bf{p}) \tag{\theequation$a$} \label{QEDkin} \\
&- e^2 \int \frac{d^3 p'}{(2 \pi)^3} \, 
\frac{1}{\left( \bf{p} - \bf{p}' \right)^2}
\bigg[ 1 - \left( \frac{p^2 + p'^2}{8 m_A^2} +
\frac{p^2 + p'^2}{8 m_B^2} \right) \tag{\theequation$b$} \label{QEDnloc} \\
&\phantom{-} \hspace{5mm} + \left( \frac{(\bf{p} \cdot \bm{\sigma}_A)
(\bf{p}' \cdot \bm{\sigma}_A)}{4 m_A^2} +
\frac{(\bf{p} \cdot \bm{\sigma}_B)
(\bf{p}' \cdot \bm{\sigma}_B)}{4 m_B^2} \right) \tag{\theequation$c$} 
\label{QEDspin} \\
&\phantom{-} \hspace{5mm} + \frac{[(\bf{p} - \bf{p}') \cdot \bm{\sigma}_A]
[(\bf{p} - \bf{p}') \cdot \bm{\sigma}_B]}{4 m_A m_B} 
- \frac{(\bf{p} - \bf{p}')^2 (\bm{\sigma}_A \cdot \bm{\sigma}_B)}
{4 m_A m_B} \tag{\theequation$d$} \label{QEDhf1} \\
&\phantom{-} \hspace{5mm} + \frac{(\bf{p} \cdot \bm{\sigma}_A)
(\bf{p}' \cdot \bm{\sigma}_A)}{2 m_A m_B} +
\frac{(\bf{p} \cdot \bm{\sigma}_B)(\bf{p}' \cdot \bm{\sigma}_B)}
{2 m_A m_B} \tag{\theequation$e$} \label{QEDhf2} \\
&\phantom{-} \hspace{5mm} + \frac{1}{4 m_A m_B} \left( (\bf{p} - \bf{p}')^2
- \frac{(p^2 - p^{\prime 2})^2}{\left( \bf{p} - \bf{p}' \right)^2}
\right) \bigg] \phi (\bf{p}') = (E - m_A - m_B) \phi (\bf{p}) \:. 
\tag{\theequation$f$} \label{QEDhf3}
\eal
Expressions \eqref{QEDkin}, \eqref{QEDnloc}, and \eqref{QEDspin} are
equal to \eqref{Ykin}, \eqref{Ynloc}, and \eqref{Yspin}, except for the
opposite sign of \eqref{QEDspin} compared to \eqref{Yspin}. However,
as discussed above, there are no contributions in the QED case analogous to
\eqref{Yreta} and \eqref{Yreth} from retardation, while in Yukawa theory 
there are no terms analogous to \eqref{QEDhf1}, \eqref{QEDhf2}, and 
\eqref{QEDhf3}. It is interesting to remark that the latter terms tend to
zero in the one-body limit $m_B \to \infty$, so that the lowest-order
fine structure calculation in the one-body limit is much simpler for the
(well-known) QED case than for Yukawa theory.

Now, if we consider the effective Hamiltonian $H_W$ instead of $H_B$,
according to the results of the previous section and Eq.\ \eqref{retardh},
the only change in the effective Schr\"odinger equations is the absence 
of the antihermitian terms \eqref{WCreta} for the Wick-Cutkosky model
and \eqref{Yreta} for Yukawa theory, while the effective Schr\"odinger
equation for QED is left unchanged as mentioned before.

\subsection{Perturbation theory for the hermitian terms}

As far as the absolute order of the correction terms is concerned, we note
the well-known fact that $p/m_r = \cal{O} (\alpha)$ in the nonrelativistic 
limit (from the expectation value of $p^2$), where $\alpha$ is the
fine structure constant. Although we will use the symbol $\alpha$
indistinctly, it is defined differently in the three theories we
consider: we take
\be
\alpha = \frac{g^2}{16 \pi m_A m_B}
\ee
in the Wick-Cutkosky model [compare with Eq.\ \eqref{WCnloc}],
\be
\alpha = \frac{g^2}{4 \pi}
\ee
in Yukawa theory, and, as usual,
\be
\alpha = \frac{e^2}{4 \pi}
\ee
in QED. Note that $p/m_r = \cal{O} (\alpha)$ is consistent with the
nonrelativistic result for the binding energy
\be
E - m_A - m_B = \cal{O} (m_r \alpha^2)
\ee 
by naive (small-$p$) power counting when we count powers of $p$ and
$p'$ indistinctly and take the integration measure $d^3 p'$ into
consideration. This kind of momentum power counting which we will refer 
to as ``IR power counting'', will be used extensively in Appendix
\ref{FWtrans}. By the same IR power counting, the perturbative corrections 
contribute to the order $m_r \alpha^4$ [the contributions from Eqs.\
\eqref{QEDhf1}, \eqref{QEDhf2}, and \eqref{QEDhf3} are suppressed by
a factor $m_A/m_B$ for $m_A \ll m_B$], except for the antihermitian terms
\eqref{WCreta} and \eqref{Yreta} which are formally of $\cal{O} (m_r
\alpha^3)$. However, as mentioned before, the contributions of the
antihermitian terms to this order turn out to vanish and they rather
contribute to $\cal{O} (m_r \alpha^4)$ like the hermitian terms.

For the rest of this section, we will discuss the application of
Rayleigh-Schr\"odinger perturbation theory in order to obtain analytical
results for the fine and hyperfine structure in the different theories.
In view of future applications with higher-order perturbative contributions,
we will perform all calculations directly in momentum space, and not
in position space like in the traditional QED fine structure calculations.
For all hermitian contributions, we apply Rayleigh-Schr\"odinger
perturbation theory to first order which is quite straightforward.
The nonrelativistic wave functions in momentum space needed for the
calculations are listed in Appendix \ref{formul} for completeness.
The matrix elements in these angular momentum eigenstates are determined
by decomposing the perturbations $H_{\text{pert}}$ in partial waves and
applying the well-known spherical harmonics addition theorem,
\be
H_{\text{pert}} (p, p', \cos \theta) = \sum_{l=0}^\infty \frac{4 \pi}{2l + 1} 
\, a_l (p, p') \sum_{m=-l}^l Y_{lm} (\hat{\bf{p}}) Y_{lm}^\ast (\hat{\bf{p}}')
\:, \label{addtheorem}
\ee
where $\hat{\bf{p}} \equiv \bf{p}/p$, $\theta$ is the angle between $\bf{p}$ 
and $\bf{p}'$, and
\be
a_l (p, p') = \frac{2l + 1}{2} \int_{-1}^1 d \cos \theta \, P_l (\cos \theta)
H_{\text{pert}} (p, p', \cos \theta) \:. \label{pwcoeff}
\ee
The explicit expressions for the relevant coefficient functions $a_l (p, p')$ 
are easily determined. For the perturbations that involve
the Pauli matrices, the formulae developed in Ref.\ \cite{WL05} 
for the application of the helicity operators $(\hat{\bf{p}} \cdot 
\bm{\sigma})$
to total angular momentum eigenstates (including the spins of the two 
fermionic constituents) are used. For convenience, they are also reproduced
in Appendix \ref{formul}. The remaining integrals over $p$ and $p'$ are
elementary, with the exception of the integrals
\be
\int_0^\infty d p \, \frac{p^{2k+1}}{(m^2 \alpha^2 + n^2 p^2)^{n+1}} 
\ln \left( \frac{p + p'}{|p - p'|} \right) \:,
\ee
where $n=1,2,3,\ldots$ and $0 \le k \le n$. All these integrals
can be obtained by differentiation with respect to $\lambda^2$ and
trivial algebraic manipulations of the fraction from 
\be
\int_0^\infty dp \, \frac{p}{\lambda^2 + p^2} \ln \left(
\frac{p + p'}{|p - p'|} \right) = \pi \arctan (p'/\lambda) \:. \label{lnint}
\ee
Equation \eqref{lnint} is derived in Appendix \ref{formul} with the help
of complex contour integration.

\subsection{Perturbation theory for the antihermitian term \label{PTat}}

Let us now develop the analogue of Rayleigh-Schr\"odinger perturbation
theory for antihermitian perturbations as they appear in the Wick-Cutkosky
model and Yukawa theory. Consider the general case of a Hamiltonian of
the form $H = H_0 + H_a$ with $H_0$ hermitian and $H_a$ antihermitian.
The eigenstates and eigenvalues of $H_0$ are supposed to be known,
\be
H_0 | \phi_n^{\ssc (0)} \rangle = E_n^{\ssc (0)} | \phi_n^{\ssc (0)} 
\rangle \:. \label{schr0}
\ee
Let us assume for simplicity that the eigenvalues (or at least the one
considered) are not degenerate. This is not true in our case, however,
the submatrices of $H_a$ in the degenerate subspaces are diagonal and
the formulae developed in the following apply despite the degeneracy.

We expand the eigenstates and eigenvalues, as usual, around the ones
of $H_0$,
\bal
| \phi_n \rangle &= | \phi_n^{\ssc (0)} \rangle + | \phi_n^{\ssc (1)} \rangle
+ \ldots \n \\
E_n &= E_n^{\ssc (0)} + E_n^{\ssc (1)} + E_n^{\ssc (2)} + \ldots \:.
\eal
Since the full Hamiltonian is not hermitian, we cannot \emph{a priori} 
assume that the eigenvalues are real nor that the eigenstates are orthogonal,
and the same is hence true for the correction terms in the expansions above.
However, we will insist on the normalization of the eigenstates and also adopt 
the usual phase convention
\be
\langle \phi_n^{\ssc (0)} | \phi_n^{\ssc (i)} \rangle \text{ real} \:.
\ee
For $i=1$, the above implies that
\be
\langle \phi_n^{\ssc (0)} | \phi_n^{\ssc (1)} \rangle = 0 \:.
\label{phasec}
\ee

Inserting the expansions into the Schr\"odinger equation for $H$, one has
the infinite tower of equations (beginning with \eqref{schr0})
\bal
H_0 | \phi_n^{\ssc (1)} \rangle + H_a | \phi_n^{\ssc (0)} \rangle 
&= E_n^{\ssc (0)} | \phi_n^{\ssc (1)} \rangle + E_n^{\ssc (1)} 
| \phi_n^{\ssc (0)} \rangle \:, \label{phin1a} \\
H_0 | \phi_n^{\ssc (2)} \rangle + H_a | \phi_n^{\ssc (1)} \rangle 
&= E_n^{\ssc (0)} | \phi_n^{\ssc (2)} \rangle + E_n^{\ssc (1)} 
| \phi_n^{\ssc (1)} \rangle + E_n^{\ssc (2)} 
| \phi_n^{\ssc (0)} \rangle \:, \label{phin2a}
\eal
etc. Projecting with $\langle \phi_n^{\ssc (0)} |$ on the first of these
equations leads to the analogue of the well-known result in hermitian 
perturbation theory,
\be
E_n^{\ssc (1)} = \langle \phi_n^{\ssc (0)} | H_a | \phi_n^{\ssc (0)} \rangle 
= - \langle \phi_n^{\ssc (0)} | H_a | \phi_n^{\ssc (0)} \rangle^\ast \:,
\label{En1a}
\ee
where we have explicitly used the antihermiticity of $H_a$ in the second 
equality. 

In our case, $H_a$ in Eqs.\ \eqref{WCreta} and \eqref{Yreta} conserves spin 
and orbital angular momentum separately, hence the angular 
and spin dependence of the wave functions $\phi_n^{\ssc (0)}$ is unchanged
under $H_a$, and the diagonal matrix elements reduce to integrals over the 
moduli of momenta. Since the operator $H_a$ and the radial (zero--order) 
wave functions are real, Eq.\ \eqref{En1a} implies that 
$E_n^{\ssc (1)} = 0$, i.e., the antihermitian perturbations give \emph{no}
contribution in first-order perturbation theory. However, $H_a$ is of 
order $p/m$ relative to the leading term in the nonrelativistic limit, while 
all the hermitian perturbations are of order $(p/m)^2$, hence the
contributions of $H_a$ in second-order perturbation theory are potentially
of the same order in $\alpha$ as the contributions of the hermitian 
perturbations in first-order perturbation theory.

To determine the second-order contributions of $H_a$, 
we use Eq.\ \eqref{phin1a} again, but now project with
$\langle \phi_m^{\ssc (0)} |$, $m \neq n$, to find an expression for
$\langle \phi_m^{\ssc (0)} | \phi_n^{\ssc (1)} \rangle$, and using the
completeness relation and the phase convention \eqref{phasec} one finds
again for the first corrections to the eigenstates the analogue of the 
result in hermitian perturbation theory,
\be
| \phi_n^{\ssc (1)} \rangle = \sum_{m \neq n} | \phi_m^{\ssc (0)} \rangle 
\frac{\langle \phi_m^{\ssc (0)} | H_a | \phi_n^{\ssc (0)} \rangle}
{E_n^{\ssc (0)} - E_m^{\ssc (0)}} \:, \label{phin1exp}
\ee
where the sum runs over all zero--order eigenstates 
$| \phi_m^{\ssc (0)} \rangle$ with $m \neq n$.

Finally, project with $\langle \phi_n^{\ssc (0)} |$ on Eq.\ \eqref{phin2a}
to find
\be
E_n^{\ssc (2)} = \langle \phi_n^{\ssc (0)} | H_a | \phi_n^{\ssc (1)} \rangle
\:, \label{En2a}
\ee
from where it follows by use of Eq.\ \eqref{phin1exp} that
\be
E_n^{\ssc (2)} = \sum_{m \neq n} \frac{\langle \phi_n^{\ssc (0)} | H_a 
| \phi_m^{\ssc (0)} \rangle \langle \phi_m^{\ssc (0)} | H_a
| \phi_n^{\ssc (0)} \rangle}{E_n^{\ssc (0)} - E_m^{\ssc (0)}}
= - \sum_{m \neq n} \frac{\abs{\langle \phi_n^{\ssc (0)} | H_a 
| \phi_m^{\ssc (0)} \rangle}^2}{E_n^{\ssc (0)} - E_m^{\ssc (0)}} \:.
\label{En2exp}
\ee
The sign appearing in the last step stems, of course, from the
antihermiticity of $H_a$. It implies, in particular, for the correction to
the ground state energy, $E_0^{\ssc (0)} < E_m^{\ssc (0)}$ for all $m \neq 0$,
that $E_0^{\ssc (2)} > 0$, contrary to the case of a hermitian perturbation.
Note that $E_n^{\ssc (2)}$ turns out to be formally of 
$\cal{O} (m_r \alpha^4)$, as anticipated, if we take 
$\langle \phi_m^{\ssc (0)} | H_a | \phi_n^{\ssc (0)} \rangle = \cal{O}
(m_r \alpha^3)$ and $E_m^{\ssc (0)} = \cal{O} (m_r \alpha^2)$.

Now the sum in Eq.\ \eqref{En2exp} is not always simple to evaluate
analytically. We will procede here in analogy with the method of
Ref.\ \cite{DL55}. The idea is to use Eq.\ \eqref{phin1a} directly to 
determine $| \phi_n^{\ssc (1)} \rangle$ rather than employing the expansion
\eqref{phin1exp}. The correction $E_n^{\ssc (2)}$ is then found from Eq.\
\eqref{En2a}. In our case, Eq.\ \eqref{phin1a} for $| \phi_n^{\ssc (1)} 
\rangle$ takes a particularly simple form due to the fact that $E_n^{\ssc (1)} 
= 0$, namely,
\be
H_0 | \phi_n^{\ssc (1)} \rangle  = E_n^{\ssc (0)} | \phi_n^{\ssc (1)} \rangle
- H_a | \phi_n^{\ssc (0)} \rangle \:, \label{phin1b}
\ee
which is the Schr\"odinger equation for the eigenvalue $E_n^{\ssc (0)}$
(with solution $| \phi_n^{\ssc (0)} \rangle$) with an additional 
inhomogeneous term. What follows is the only part of the calculation that
we have performed in position space, for reasons to become clear shortly.
It is not difficult, of course, to transform every step of the argument
to momentum space. We will also present a different approach to the
calculation of $E_n^{\ssc (2)}$ in Appendix \ref{FWtrans} which proceeds
entirely in momentum space.

To proceed with the calculation in position space, we then need the 
explicit expression for $H_a$ in position space. To this end, start with
\be
\frac{1}{\abs{\bf{p} - \bf{p}'}} = \frac{1}{2 \pi^2} \int d^3 r \, 
e^{- i (\bf{p} - \bf{p}') \cdot \bf{r}} \frac{1}{r^2} \:. \label{coulf}
\ee
Taking the derivate of Eq.\ \eqref{coulf} with respect to $\bf{p}$ and 
multiplying with $(\bf{p} + \bf{p}')$ gives
\be
\frac{p^2 - p'^2}{\abs{\bf{p} - \bf{p}'}^3} = \frac{1}{2 \pi^2} \int d^3 r \,
\frac{\bf{r}}{r^2} \cdot \left( e^{-i \bf{p} \cdot \bf{r}} 
\frac{\d}{\d \bf{r}} e^{i \bf{p}' \cdot \bf{r}} - e^{i \bf{p}' \cdot \bf{r}} 
\frac{\d}{\d \bf{r}} e^{-i \bf{p} \cdot \bf{r}} \right) \:,
\ee
so that
\bal
\lefteqn{H_a \phi_n^{\ssc (0)} (\bf{r})} \n \\
&= \int \frac{d^3 p}{(2 \pi)^3} \, 
e^{i \bf{p} \cdot \bf{r}} \int \frac{d^3 p'}{(2 \pi)^3} \,
\frac{\alpha}{2 \pi m_r} \int d^3 r' \, \frac{\bf{r}'}{r'^2} \cdot \left( 
e^{-i \bf{p} \cdot \bf{r}'} \frac{\d}{\d \bf{r}'} e^{i \bf{p}' \cdot \bf{r}'} 
- e^{i \bf{p}' \cdot \bf{r}'} \frac{\d}{\d \bf{r}'} 
e^{-i \bf{p} \cdot \bf{r}'} \right) \phi_n^{\ssc (0)} (\bf{p}') \:.
\eal
Integrating over $\bf{p}$, $\bf{p}'$, and $\bf{r}'$ finally leads to
\be
H_a \phi_n^{\ssc (0)} (\bf{r}) 
= \frac{\alpha}{2 \pi m_r} \left( \frac{2}{r} \frac{\d}{\d r} +
\frac{1}{r^2} \right) \phi_n^{\ssc (0)} (\bf{r}) \:.
\ee

For the solution of Eq.\ \eqref{phin1b}, we now make the ansatz
\be
\phi_{n l m}^{\ssc (1)} ({\bf r}) = \chi_{n l} (r) \phi_{n l m}^{\ssc (0)} 
({\bf r}) \:, \label{phi1ans}
\ee
where
\be
\phi_{n l m}^{\ssc (0)} ({\bf r}) = \frac{u_{n l}^{\ssc (0)} (r)}{r} 
Y_{l m} (\hat{\bf{r}}) 
\ee
is a solution of Eq.\ \eqref{schr0}, and still has to be 
to be multiplied with the appropriate Pauli spinors to describe the spin
orientations of fermions $A$ and $B$ in the case of Yukawa theory.
When we insert this ansatz into Eq.\ \eqref{phin1b} and use the Schr\"odinger
equation \eqref{schr0} for $\phi_{n l m}^{\ssc (0)} ({\bf r})$, the equation
\be
\left( u_{n l}^{\ssc (0)} (r) \frac{d^2}{d r^2} + 2 \frac{d u_{n l}^{\ssc (0)}}
{d r} \frac{d}{d r} \right) \chi_{n l} (r) 
= \frac{\alpha}{\pi} \left( \frac{2}{r} \frac{d}{d r} 
- \frac{1}{r^2} \right) u_{n l}^{\ssc (0)} (r) \label{defchi}
\ee
for $\chi_{n l} (r)$ results. The terms on the left-hand-side of Eq.\ 
\eqref{defchi} stem from the second $r$-derivative in the kinetic term of 
$H_0$.

Equation \eqref{defchi} is identically fulfilled, i.e., independently of the
function $u_{n l}^{\ssc (0)} (r)$, for
\be
\frac{d \chi_{n l}}{d r} = \frac{\alpha}{\pi r} \:. \label{diffchi}
\ee
Being a linear differential equation for $d \chi/d r$, Eq.\ \eqref{defchi}
has a one-dimensional (affine) vector space of solutions. We will hence
have a look at the solutions of the corresponding homogeneous equation.
Neglecting powers of $r$, $u_{n l}^{\ssc (0)} (r)$ falls to zero for large
$r$ like
\be
u_{n l}^{\ssc (0)} (r) \sim e^{-r/n a_0} \:,
\ee
$a_0$ being the Bohr radius. It follows that the nontrivial solutions of the
homogeneous equation behave like
\be
\frac{d \chi_{n l}}{d r} \sim e^{2 r/n a_0}
\ee
for large $r$, and the corresponding functions
$\phi_{n l m}^{\ssc (1)} ({\bf r})$ in Eq.\ \eqref{phi1ans} are not
acceptable as bound state solutions. As a result, we have to take the
trivial solution of the homogeneous equation, and Eq.\ \eqref{diffchi} is
in fact the physical solution of Eq.\ \eqref{defchi}.

The simplicity of the ansatz \eqref{phi1ans} which leads to the solution
\eqref{diffchi}, is the reason for our present use of position space.
In momentum space, Eq.\ \eqref{phi1ans} corresponds to a convolution 
which makes the latter formulation a little less transparent in
the solution of Eq.\ \eqref{phin1b}.

From the integration of Eq.\ \eqref{diffchi} we obtain
\be
\chi_{n l} (r) = \frac{\alpha}{\pi} \ln \frac{2 r}{n a_0} + C_{n l} \:, 
\label{intchi}
\ee
with a convenient normalization of the argument of the logarithm. The
integration constant $C_{n l}$ is fixed by the phase convention
\eqref{phasec} which is where the $(n,l)$-dependence of $\chi_{n l} (r)$
enters [compare with Eq.\ \eqref{diffchi}]. The constant actually plays no 
role at all in the calculation of the perturbed energy with the help of Eq.\ 
\eqref{En2a}, given that $\langle \phi_n^{\ssc (0)} | H_a | \phi_n^{\ssc (0)} 
\rangle = 0$ as discussed after Eq.\ \eqref{En1a}.

The constant $C_{n l}$ is important, however, when one is interested in 
the wave functions themselves. The integrals corresponding to Eq.\
\eqref{phasec} can be evaluated with the help of integral tables,
for example, Ref.\ \cite{GR00}. The explicit results for the lowest-lying 
states are
\bal
C_{10} &= \frac{\alpha}{\pi} \left( \gamma_E - \frac{3}{2} \right) \, \n \\
C_{20} &= \frac{\alpha}{\pi} \left( \gamma_E - \frac{9}{4} \right) \:, \n \\
C_{21} &= \frac{\alpha}{\pi} \left( \gamma_E - \frac{25}{12} \right) \:,
\eal
with the Euler-Mascheroni constant $\gamma_E$. It is tempting to
speculate that the logarithms of $r$ in the corrections to the wavefunctions,
when summed up to all orders of the perturbative expansion, combine into
an $\alpha$-dependent power of $r$. This gives a hint to a possible
improvement in the choice of the basis functions used for the numerical
solution of Eqs.\ \eqref{schrWC} and \eqref{schrY}.

From Eqs.\ \eqref{En2a}, \eqref{phi1ans},  and \eqref{intchi} we can now 
calulate the corrections $E_n^{\ssc (2)}$ to the energy, again with the help 
of integral tables. However, there is a different, interesting, and somewhat 
simpler way to calculate these corrections. To get there, we manipulate the
expression \eqref{En2a} for $E_n^{\ssc (2)}$ in the following way:
\bal
E_n^{\ssc (2)} &= \frac{\alpha}{2 \pi m_r} \int_0^\infty dr \, r^2 \, 
\frac{u_n^{\ssc (0)} (r)} {r} \left( \frac{2}{r} \frac{d}{d r} + 
\frac{1}{r^2} \right) \frac{\chi (r) u_n^{\ssc (0)} (r)}{r} \n \\
&= \frac{\alpha}{2 \pi m_r} \int_0^\infty dr \, r^2 \, 
\frac{u_n^{\ssc (0)} (r)}{r} \left( \frac{2}{r} \frac{d \chi}{d r} \right) 
\frac{u_n^{\ssc (0)} (r)}{r} \n \\
&\phantom{=} \hspace{1cm} {}
+ \frac{\alpha}{2 \pi m_r} \int_0^\infty dr \, r^2 
\, \frac{\chi (r) u_n^{\ssc (0)} (r)}{r} 
\left( \frac{2}{r} \frac{d}{d r} + \frac{1}{r^2} \right) 
\frac{u_n^{\ssc (0)} (r)}{r} \n \\
&= \frac{\alpha}{2 \pi m_r} \langle \phi_n^{\ssc (0)} | \, \frac{2 \alpha}
{\pi r^2} \, | \phi_n^{\ssc (0)} \rangle
- \frac{\alpha}{2 \pi m_r} \int_0^\infty dr \, 
r^2 \, \frac{u_n^{\ssc (0)} (r)}{r} \left( \frac{2}{r} \frac{d}{d r} + 
\frac{1}{r^2} \right) \frac{\chi (r) u_n^{\ssc (0)} (r)}{r} \:,
\label{Eanti2}
\eal
where we have used Eq.\ \eqref{diffchi} in the last step. It follows that
\be
E_n^{\ssc (2)} = \frac{\alpha^2}{2 \pi^2 m_r} \, \langle \phi_n^{\ssc (0)} | 
\, \frac{1}{r^2} \, | \phi_n^{\ssc (0)} \rangle \:, \label{Eantiexp}
\ee
which has the form of a first-order correction for the hermitian perturbation
\be
\frac{\alpha^2}{2 \pi^2 m_r} \, \frac{1}{r^2} \:, \label{Eantiop}
\ee
or, equivalently,
\be
\frac{\alpha^2}{m_r} \, \frac{1}{\abs{\bf{p} - \bf{p}'}} \label{Eantiopm}
\ee
in momentum space. Expression \eqref{Eantiexp} can then be evaluated by
the standard methods desribed in the previous subsection.

Let us emphasize again that the contribution \eqref{Eantiexp} is due to
the retardation of the interaction in the Wick-Cutkosky model and
Yukawa theory, and that it is absent in the hermitian Okubo Hamiltonian.
This situation is reminiscent of a discussion in the older literature on 
the effective one-boson exchange (OBE) description of the nucleon-nucleon 
interaction: the Gross equation develops a repulsive contribution to the 
interaction potential in a (not completely systematic) expansion around the 
nonrelativistic limit, while such a contribution is absent in the 
Blankenbecler-Sugar-Logunov-Tavkhelidze (BSLT) equation \cite{Gro74}. 
Incidentally, the Gross equation includes retardation in its kernel while 
the BSLT equation does not (to lowest order).
In Ref.\ \cite{Gro74}, the repulsive contribution to the potential was
associated with the repulsive core of the nucleon-nucleon interaction.
Since the exchange of a scalar plays an important role in the effective
OBE description giving the dominant attractive contribution to the
intermediate-range potential, it is actually possible that the contribution
\eqref{Eantiop} is relevant to physics in the context of the OBE potential.
Another interesting remark in this respect is that the full Gross
equation is plagued by singularities which complicate its numerical
solution \cite{GVH92}, while there is no such problem with Eq.\ \eqref{schrY}
\cite{WL05}.

How is it possible that the two effective Hamiltonians $H_B$ and $H_W$
lead to different predictions about the lowest-order fine and hyperfine
structure? Both effective Hamiltonians have been calculated to order
$H_1^2$ or $g^2$, and the next terms in the perturbative expansion are
of the order $g^4$. It is possible that these latter terms
contribute to order $m_r \alpha^4$, i.e., to the lowest-order fine
and hyperfine structure, and these contributions need not be
the same in the two cases. Summing all contributions to order $m_r \alpha^4$,
from the order-$g^2$ and order-$g^4$ terms of the effective Hamiltonian,
however, must lead to the same (the complete) result for the lowest-order
fine and hyperfine structure. It is then also clear that we cannot be
sure that the terms of order $\alpha^4$ obtained in the present paper
represent the complete fine or hyperfine structure in any of the two
cases. In fact, the difference between the results for $H_B$ and $H_W$
shows that at least one of them (or both) must be incomplete. From the
results of Appendix \ref{FWtrans} it appears plausible that the
contribution \eqref{Eantiop} is obtained from the order-$g^4$ term
of the effective Hamiltonian $H_W$, so that at least in the latter case
the complete fine and hyperfine structure probably cannot be obtained from the
$g^2$-term alone. On the other hand, the $g^2$-term is sufficient to
generate the complete fine and hyperfine structure in the case of QED
(in the Coulomb gauge), for both $H_B$ and $H_W$, as we will show in the
next section. In particular, then, there really are cases where the 
$g^2$-term is sufficient for our purpose. The situation remains
unclear in the Wick-Cutkosky model and Yukawa theory until we have
determined all terms of the order $g^4$ in the effective Hamiltonian,
although it might well be that the $g^2$-term in $H_B$ (and thus not in
$H_W$) generate the lowest-order fine and hyperfine structure completely.
In fact, it may be a reasonable criterium for a ``good'' relativistic
bound state equation that the lowest-order approximation produce all
the lowest-order fine and hyperfine structure.

\section{Results and discussion}

We will now give explicit results for the different contributions to the
fine and hyperfine stucture in the three theories considered, for the
states with principal quantum numbers $n=1,2$. These results will be
discussed with respect to consistency, comparison with the numerical
solutions of the full effective Schr\"odinger equations and with other
relativistic bound state equations. In the case of QED, the well-known
results will be reproduced completely, while the results for the
lowest-order fine and hyperfine structure in the Wick-Cutkosky model
and Yukawa theory have, to the best of our knowledge, not been obtained 
before.

\subsection{Wick-Cutkosky model}

By applying the methods detailed in the previous section (and in the
appendices) to the lowest-lying states, we arrive at the explicit results
for the relativistic energy corrections $\Delta E (n L)$ ($L = S, P, D, 
\ldots$) cited below. The different contributions specified follow the order 
in the expanded effective Hamiltonian $H_B$, i.e., from left to right we have 
the  relativistic correction to the kinetic energy in Eq.\ \eqref{WCkin}, the
correction term in Eq.\ \eqref{WCnloc}, the antihermitian term \eqref{WCreta},
and the term on the left-hand side of Eq.\ \eqref{WCreth}. The corresponding
results for $H_W$ are obtained by just leaving out the contribution from
the antihermitian term (third term in every result). The explicit 
expressions are
\bal
\Delta E (1 S) &= \left[ - \frac{5}{8} \left( \frac{1}{m_A^3} + 
\frac{1}{m_B^3} \right) m_r^3 + \frac{3}{2} \left( \frac{1}{m_A^2} 
+ \frac{1}{m_B^2} \right) m_r^2 \right. \n \\
&\phantom{=} \hspace{3cm} \left. + \frac{1}{\pi^2}
- \frac{1}{2} \left( \frac{1}{m_A^2} + \frac{1}{m_B^2} \right) m_r^2 
\right] m_r \alpha^4 \:, \n \\[2mm]
\Delta E (2 S) &= \left[ - \frac{13}{128} \left( \frac{1}{m_A^3} + 
\frac{1}{m_B^3} \right) m_r^3 + \frac{7}{32} \left( \frac{1}{m_A^2} 
+ \frac{1}{m_B^2} \right) m_r^2 \right. \n \\
&\phantom{=} \hspace{3cm} \left. + \frac{1}{8 \pi^2}
- \frac{1}{16} \left( \frac{1}{m_A^2} + \frac{1}{m_B^2} \right) m_r^2 
\right] m_r \alpha^4 \:, \n \\[2mm]
\Delta E (2 P) &= \left[ - \frac{7}{384} \left( \frac{1}{m_A^3} + 
\frac{1}{m_B^3} \right) m_r^3 + \frac{5}{96} \left( \frac{1}{m_A^2} 
+ \frac{1}{m_B^2} \right) m_r^2 \right. \n \\
&\phantom{=} \hspace{3cm} \left. + \frac{1}{24 \pi^2}
- \frac{1}{48} \left( \frac{1}{m_A^2} + \frac{1}{m_B^2} \right) m_r^2 
\right] m_r \alpha^4 \:. \label{WCres}
\eal
The most important general features of these results are the following:
the sum of all corrections is positive for any of these levels, and for
any mass ratio $m_A/m_B$. After taking into account these corrections,
the energy of the $2 S$ level lies \emph{above} the $2 P$ level (these
levels are, of course, degenerate in the nonrelativistic limit). Both
of these features are due to the dominant correction term from Eq.\ 
\eqref{WCnloc}. This term arises from the expansion of the inverse
square roots of kinetic energies in Eqs.\ \eqref{HBWC} or \eqref{schrWC},
which are characteristic of the nonlocalizability of point particles in
relativistic quantum theories. The overall contribution of retardation,
corrections \eqref{WCreta} and \eqref{WCreth}, is negative, hence the
retardation of the potential has an attractive effect (see, however,
the discussion of the antihermitian term at the end of Subsection
\ref{PTat}). Incidentally, the relation between the energy corrections
from these two terms is the same for all energy levels.

\begin{figure}
\begin{center}
\resizebox{10cm}{!}{\includegraphics*{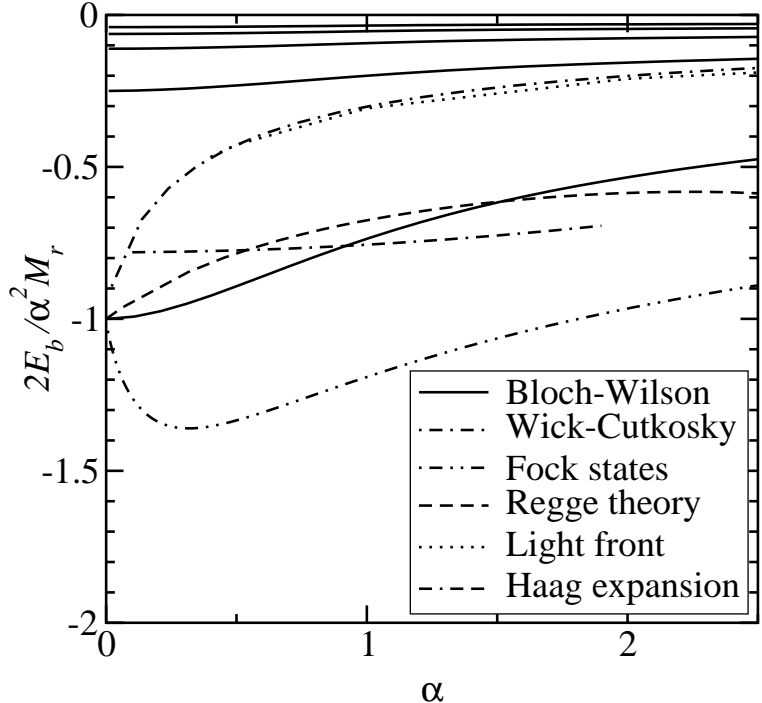}}
\end{center}
\caption{\label{figWCq} The spectrum of binding energies 
$E_b = E - m_A - m_B$ for $S$--states in
the equal--mass case, compared to the ground state energies of
the Wick-Cutkosky model \cite{WC54}, the Hamiltonian eigenvalue 
equation in a Fock space truncation \cite{LB01}, the Regge theory predictions 
\cite{WLS00}, the light--front calculation \cite{MC00}, and the Haag 
expansion results \cite{GRS95} in their domain of validity.}
\end{figure}

The most important property of all, however, is probably the fact that the
corrections to the energy are of order $m_r \alpha^4$, as one might have
expected from the well-known fine and hyperfine structure of hydrogen.
A look at Fig.\ \ref{figWCq} which we have reproduced here, for convenience,
from Ref.\ \cite{WL02}, shows that this feature is nicely reproduced
by the results from a numerical solution of the full effective 
Schr\"odinger equation \eqref{schrWC} \cite{WL02}. Observe that the
energy eigenvalues presented in Fig.\ \ref{figWCq} are normalized to
the nonrelativistic inonization energy $m_r \alpha^2/2$, hence the
order $m_r \alpha^4$ of the relativistic corrections corresponds to a
quadratic behavior of the curve near $\alpha = 0$. 

It is a remarkable fact that the numerical curves for all other relativistic 
bound state equations presented in Fig.\ \ref{figWCq} do \emph{not}
show a quadratic behavior for small $\alpha$ (disregarding the results from
the Haag expansion which does not have a proper nonrelativistic limit),
which points to the existence of an order-$(m_r \alpha^3)$ term in their
expansion around the nonrelativistic limit. In fact, it is known
that there appears a ``curious'' term of order $m_r \alpha^3 \ln \alpha$
in the ladder approximation to the Bethe-Salpeter equation in such an 
expansion \cite{FH73}. A simple power-counting argument cannot exclude 
the existence of contributions to the order $m_r \alpha^3$ (or $m_r \alpha^3
\ln \alpha$) from the terms of order $H_1^4$ in the
expansion of $H_B$ derived from the Dyson series \eqref{dyson} which are not
considered in the present paper. However, from our experience with QED
(e.g., the hydrogen atom and positronium) including terms of the order
$H_1^4$, such contributions are certainly not expected. Note that a 
contribution of 
the next-higher order $m_r \alpha^4$ from the order-$H_1^4$ terms is
necessary for the consistence of the results for $H_B$ and $H_W$ presented
here which differ by a contribution of the order $m_r \alpha^4$, as we have 
seen (compare with the discussion at the end of Subsection \ref{PTat}).
The definite solution of these issues, most importantly the question of
whether the results presented here for the fine structure (from $H_B$)
are complete, has to await the (complicated) evaluation of the $\cal{O}
(H_1^4)$ terms by the present or a different method.

\begin{figure}
\begin{center}
\begin{picture}(110,79)
\put(10,-39){\psfig{figure=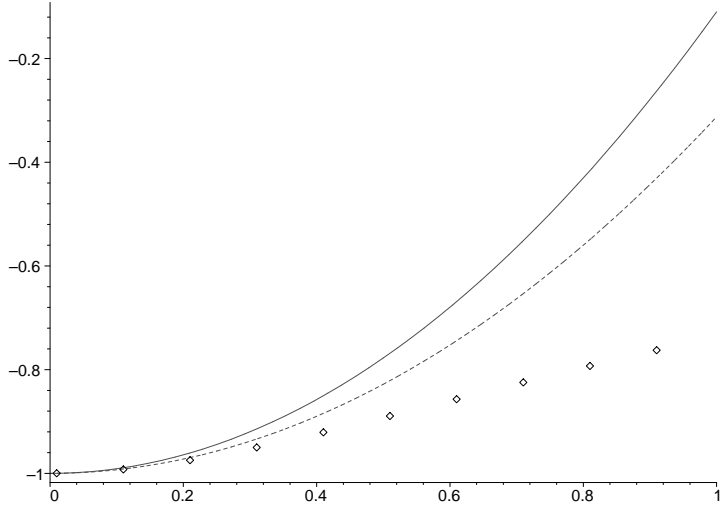,width=80mm}}
\end{picture}
\end{center}
\caption{\label{figWCn} The numerical results (diamonds) for the ground state 
binding energy, normalized as in Fig.\ \ref{figWCq}, compared with the
perturbative results \eqref{WCres} (solid line). We have also plotted the
perturbative results without the contribution from the antihermitian term
(dashed line).}
\end{figure}

To give a better idea of how the numerical results relate to the perturbative
expansion around the nonrelativistic limit, we have plotted both in Fig.\
\ref{figWCn} for the ground state. The approximation by the perturbative 
expression is good only way below $\alpha = 0.1$. We have also plotted the 
curve corresponding to all hermitian terms, i.e., omitting the contribution 
from Eq.\ \eqref{WCreta} in Eq.\ \eqref{WCres}. Note that the latter curve
lies closer to the numerical data points than the curve for the complete
expressions, for intermediate values of $\alpha$.

\subsection{Yukawa theory}

For Yukawa theory, the calculation is technically, but not essentially,
more complicated than for the Wick-Cutkosky model, due to the spin of the
fermions. The necessary techniques for the evaluation of the relativistic
corrections are presented in Appendix \ref{formul}. We will label the
states by the standard spectroscopic notation $n \supfi{2 S + 1} L_J$.
The explicit results for principal qunatum numbers $n = 1,2$ are, again
term by term in the order they appear in the momentum expansion of the
effective Hamiltonian, Eqs.\ \eqref{Ykin}--\eqref{Yspin},
\bal
\Delta E (1 \supfi{1} S_0) &= \Delta E (1 \supfi{3} S_1) 
= \left[ - \frac{5}{8} \left( \frac{1}{m_A^3} + \frac{1}{m_B^3} \right) m_r^3 
+ \frac{3}{4} \left( \frac{1}{m_A^2} + \frac{1}{m_B^2} \right) m_r^2 \right. 
\n \\
&\phantom{=} \hspace{3cm} \left. + \frac{1}{\pi^2} + \left( - \frac{1}{2} 
+ \frac{1}{4} \right) \left( \frac{1}{m_A^2} + \frac{1}{m_B^2} \right) m_r^2 
\right] m_r \alpha^4 \:, \n \\[2mm]
\Delta E (2 \supfi{1} S_0) &= \Delta E (2 \supfi{3} S_1) 
=\left[ - \frac{13}{128} \left( \frac{1}{m_A^3} + 
\frac{1}{m_B^3} \right) m_r^3 + \frac{7}{64} \left( \frac{1}{m_A^2} 
+ \frac{1}{m_B^2} \right) m_r^2 \right. \n \\
&\phantom{=} \hspace{3cm} \left. + \frac{1}{8 \pi^2} + \left(
- \frac{1}{16} + \frac{3}{64} \right) \left( \frac{1}{m_A^2} + 
\frac{1}{m_B^2} \right) m_r^2 \right] m_r \alpha^4 \:, \n \\[2mm]
\Delta E (2 \supfi{3} P_0) &= \left[ - \frac{7}{384} \left( \frac{1}{m_A^3} + 
\frac{1}{m_B^3} \right) m_r^3 + \frac{5}{192} \left( \frac{1}{m_A^2} 
+ \frac{1}{m_B^2} \right) m_r^2 \right. \n \\
&\phantom{=} \hspace{3cm} \left. + \frac{1}{24 \pi^2} + \left( - \frac{1}{48} 
+ \frac{9}{192} \right) \left( \frac{1}{m_A^2} + \frac{1}{m_B^2} \right) m_r^2 
\right] m_r \alpha^4 \:, \n \\[2mm]
\Delta E' (2 \supfi{3} P_1) &= \left[ - \frac{7}{384} \left( \frac{1}{m_A^3} + 
\frac{1}{m_B^3} \right) m_r^3 + \frac{5}{192} \left( \frac{1}{m_A^2} 
+ \frac{1}{m_B^2} \right) m_r^2 \right. \n \\
&\phantom{=} \hspace{3cm} \left. + \frac{1}{24 \pi^2} + \left( - \frac{1}{48} 
+ \frac{7}{192} \right) \left( \frac{1}{m_A^2} + \frac{1}{m_B^2} \right) m_r^2 
\right] m_r \alpha^4 \:, \n \\[2mm]
\Delta E' (2 \supfi{1} P_1) &= \left[ - \frac{7}{384} \left( \frac{1}{m_A^3} + 
\frac{1}{m_B^3} \right) m_r^3 + \frac{5}{192} \left( \frac{1}{m_A^2} 
+ \frac{1}{m_B^2} \right) m_r^2 \right. \n \\
&\phantom{=} \hspace{3cm} \left. + \frac{1}{24 \pi^2} + \left( - \frac{1}{48} 
+ \frac{5}{192} \right) \left( \frac{1}{m_A^2} + \frac{1}{m_B^2} \right) m_r^2 
\right] m_r \alpha^4 \:, \n \\[2mm]
\Delta E (2 \supfi{3} P_2) &= \left[ - \frac{7}{384} \left( \frac{1}{m_A^3} + 
\frac{1}{m_B^3} \right) m_r^3 + \frac{5}{192} \left( \frac{1}{m_A^2} 
+ \frac{1}{m_B^2} \right) m_r^2 \right. \n \\
&\phantom{=} \hspace{3cm} \left. + \frac{1}{24 \pi^2} + \left( - \frac{1}{48} 
+ \frac{3}{192} \right) \left( \frac{1}{m_A^2} + \frac{1}{m_B^2} \right) m_r^2 
\right] m_r \alpha^4 \:. \label{Yres}
\eal
The results for the states $2 \supfi{3} P_1$ and $2 \supfi{1} P_1$ bear
primes because these states (which are degenerate in the nonrelativistic
limit) mix through the equal off-diagonal matrix elements
\be
- \frac{\sqrt{2}}{96} \left( \frac{1}{m_A^2} - \frac{1}{m_B^2} \right)
m_r^3 \alpha^4 \:. \label{Yoffd}
\ee
To obtain the energy corrections properly, one hence has to diagonalize
the corresponding $2 \times 2$ matrix with diagonal elements 
$\Delta E' (2 \supfi{3} P_1)$ and $\Delta E' (2 \supfi{1} P_1)$.
The explicit expressions for the eigenvalues are not too illuminating.

Compared to the results for the Wick-Cutkosky model, in the Yukawa case
the contributions from the ``nonlocal'' term \eqref{WCnloc} are reduced
to half their value by a contribution from the normalization of the
Dirac spinors in \eqref{Ynloc}, and a positive contribution from the
spin structure enters. The fact that the ``nonlocal'' contribution is
reduced leads to a sign change for the relativistic corrections of the
$1 S$ states depending on the mass ratio: in the case of equal masses, all
corrections are positive, while the corrections for $1 \supfi{1} S_0$
and $1 \supfi{3} S_1$ turn negative in the one-body limit $m_B \to \infty$.

As for the ``mixing'' states $2 \supfi{3} P_1$ and $2 \supfi{1} P_1$,
in the case of equal masses the off-diagonal elements \eqref{Yoffd} 
vanish, and $\Delta E' (2 \supfi{3} P_1)$ and $\Delta E' (2 \supfi{1} P_1)$
directly give the energy corrections of the states. Observe the equal
spacing between the energy corrections of the four $2 P$ states 
in this case which
is nicely reproduced by the numerical results up to intermediate coupling
constants. In the one-body limit, on the other hand, the states mix
strongly, and one of the eigenstates becomes degenerate with the state
$2 \supfi{3} P_0$, the other with $2 \supfi{3} P_2$. These degeneracies
and the ones of $1 \supfi{1} S_0$ with $1 \supfi{3} S_1$ and of 
$2 \supfi{1} S_0$ with $2 \supfi{3} S_1$, are exact in the one-body limit,
for any value of the coupling constant. The reason for these twofold
degeneracies is that the spin of fermion $B$ decouples from the dynamics in 
the limit $m_B \to \infty$. In fact, the states $2 \supfi{3} P_0$ and the
mixture of $2 \supfi{3} P_1$ and $2 \supfi{1} P_1$ degenerate with the former,
are eigenstates of the total angular momentum (relative orbital angular
momentum and spin) of fermion $A$ with eigenvalue $j_A = 1/2$, while
state $2 \supfi{3} P_2$ and the other linear combination of $2 \supfi{3} P_1$ 
and $2 \supfi{1} P_1$ are $j_A$ eigenstates with eigenvalue 3/2. Note that 
the $j_A = 1/2$ state lies higher in energy than the $j_A = 3/2$ state.

Close to the one-body limit, we can check for hyperfine structure
by expanding the expressions for the energy corrections of the $2 P$ states
in Eq.\ \eqref{Yres} and the off-diagonal elements \eqref{Yoffd} in powers of 
$m_A/m_B$. Obviously, there is no hyperfine splitting for the $1 S$ and
the $2 S$ states to any order in $m_A/m_B$ (and to order $m_r \alpha^4$). 
After diagonalizing the submatrix for the states $2 \supfi{3} P_1$ and 
$2 \supfi{1} P_1$, one finds \emph{no hyperfine splitting} to the order 
$m_A/m_B$, while there does appear such a splitting to the next order, 
$(m_A/m_B)^2$. These results on the hyperfine structure have been anticipated 
in Ref.\ \cite{WL05}.

The perturbative results \eqref{Yres} show all the qualitative features that 
we have found in the numerical solutions of the effective Schr\"odinger 
equation \eqref{schrY} in Ref.\ \cite{WL05}. In quantitative terms, the
agreement between numerical and perturbative analytical results is
similar to the case of the Wick-Cutkosky model (see Fig.\ \ref{figWCn}) for 
the $S$ states, and somewhat better for the $P$ states.
Just like in Fig.\ \ref{figWCn}, the perturbative curve
without the contribution of the antihermitian term lies closer to the
numerical results for intermediate coupling constants than the full
perturbative curve. It appears that the antihermitian term is only important
for very small $\alpha$ and does practically not contribute for larger
values of $\alpha$. The most clear-cut case is the one of 
$\Delta E (2 \supfi{1} S_0)$ = $\Delta E (2 \supfi{3} S_1)$ in the one-body
limit where the complete perturbative correction is positive, while
its hermitian part is negative. In this case, the numerical values for
very small $\alpha$ increase with the coupling constant as expected from the 
perturbative results, but then start to decrease and become negative roughly 
at $\alpha = 0.35$, loosely following the ``hermitian'' curve for 
larger $\alpha$.

The peculiar role that is played by the contributions from the
antihermitian term might lead one to the believe that these contributions
are spurious and may cancel with other contributions from terms of the
order $H_1^4$ in the expansion of the effective Hamiltonian. In this sense,
the Okubo Hamiltonian $H_W$ would be a better choice for an effective
Hamiltonian, because it simply does not contain the antihermitian term.
However, the results of Appendix \ref{FWtrans} suggest that, quite to the
contrary, the terms of the order $H_1^4$ in the expansion of $H_W$
generate the hermitian equivalent of the term in question.

\subsection{Quantum electrodynamics}

Finally, we will present our results for QED in the Coulomb gauge, i.e.,
the expanded effective Hamiltonian of Eqs.\ \eqref{QEDkin}-\eqref{QEDhf3}.
The explicit expressions are \cite{Con03}, term by term,
\bal
\Delta E (1 \supfi{1} S_0) &= 
\bigg[ - \frac{5}{8} \left( \frac{1}{m_A^3} + \frac{1}{m_B^3} \right) m_r^3 
+ \left( \frac{3}{4} - \frac{1}{4} \right) \left( \frac{1}{m_A^2} + 
\frac{1}{m_B^2} \right) m_r^2 
\n \\
&\phantom{=} \hspace{5cm} + \big( - 2 - 1 + 0 \big) 
\frac{m_r^2}{m_A m_B} \bigg] m_r \alpha^4 \:, \n \\[2mm]
\Delta E (1 \supfi{3} S_1) &= \left[ - \frac{5}{8} \left( \frac{1}{m_A^3} + 
\frac{1}{m_B^3} \right) m_r^3 + \left( \frac{3}{4} - \frac{1}{4} \right) 
\left( \frac{1}{m_A^2} + \frac{1}{m_B^2} \right) m_r^2 \right.
\n \\
&\phantom{=} \hspace{5cm} \left. + \left( \frac{2}{3} - 1 + 0 \right) 
\frac{m_r^2}{m_A m_B} \right] m_r \alpha^4 \:, \n \\[2mm]
\Delta E (2 \supfi{1} S_0) &=
\left[ - \frac{13}{128} \left( \frac{1}{m_A^3} + 
\frac{1}{m_B^3} \right) m_r^3 + \left( \frac{7}{64} - \frac{3}{64} \right) 
\left( \frac{1}{m_A^2} + \frac{1}{m_B^2} \right) m_r^2 \right. \n \\
&\phantom{=} \hspace{5cm} \left. + \left( - \frac{1}{4} - \frac{3}{16}
+ 0 \right) \frac{m_r^2}{m_A m_B} \right] m_r \alpha^4 \:, \n \\[2mm]
\Delta E (2 \supfi{3} S_1) &=\left[ - \frac{13}{128} \left( \frac{1}{m_A^3} + 
\frac{1}{m_B^3} \right) m_r^3 + \left( \frac{7}{64} - \frac{3}{64} \right) 
\left( \frac{1}{m_A^2} + \frac{1}{m_B^2} \right) m_r^2 \right. \n \\
&\phantom{=} \hspace{5cm} \left. + \left( \frac{1}{12} - \frac{3}{16}
+ 0 \right) \frac{m_r^2}{m_A m_B} \right] m_r \alpha^4 \:, \n \\[2mm]
\Delta E (2 \supfi{3} P_0) &= \left[ - \frac{7}{384} \left( \frac{1}{m_A^3} + 
\frac{1}{m_B^3} \right) m_r^3 + \left( \frac{5}{192} - \frac{3}{64} \right) 
\left( \frac{1}{m_A^2} + \frac{1}{m_B^2} \right) m_r^2 \right. \n \\
&\phantom{=} \hspace{5cm} \left. + \left( - \frac{1}{24} - \frac{3}{16}
+ \frac{1}{24} \right) \frac{m_r^2}{m_A m_B} \right] m_r \alpha^4 \:, 
\n \\[2mm]
\Delta E' (2 \supfi{3} P_1) &= \left[ - \frac{7}{384} \left( \frac{1}{m_A^3} + 
\frac{1}{m_B^3} \right) m_r^3 + \left( \frac{5}{192} - \frac{7}{192} \right)  
\left( \frac{1}{m_A^2} + \frac{1}{m_B^2} \right) m_r^2 \right. \n \\
&\phantom{=} \hspace{5cm} \left. + \left( \frac{1}{48} - \frac{7}{48} +
\frac{1}{24} \right) \frac{m_r^2}{m_A m_B} \right] m_r \alpha^4 \:, 
\n \\[2mm]
\Delta E' (2 \supfi{1} P_1) &= \left[ - \frac{7}{384} \left( \frac{1}{m_A^3} + 
\frac{1}{m_B^3} \right) m_r^3 + \left( \frac{5}{192} - \frac{5}{192} \right) 
\left( \frac{1}{m_A^2} + \frac{1}{m_B^2} \right) m_r^2 \right. \n \\
&\phantom{=} \hspace{5cm} \left. + \left( 0 - \frac{5}{48} + \frac{1}{24}
\right) \frac{m_r^2}{m_A m_B} \right] m_r \alpha^4 \:, \n \\[2mm]
\Delta E (2 \supfi{3} P_2) &= \left[ - \frac{7}{384} \left( \frac{1}{m_A^3} + 
\frac{1}{m_B^3} \right) m_r^3 + \left( \frac{5}{192} - \frac{1}{64} \right) 
\left( \frac{1}{m_A^2} + \frac{1}{m_B^2} \right) m_r^2 \right. \n \\
&\phantom{=} \hspace{5cm} \left. + \left( - \frac{1}{240} - \frac{1}{16}
+ \frac{1}{24} \right) \frac{m_r^2}{m_A m_B} \right] m_r \alpha^4 \:. 
\label{QEDres}
\eal
As in the case of Yukawa theory, the states $2 \supfi{3} P_1$ and 
$2 \supfi{1} P_1$ mix through the equal off-diagonal matrix elements
\be
\frac{\sqrt{2}}{96} \left( \frac{1}{m_A^2} - \frac{1}{m_B^2} \right)
m_r^3 \alpha^4 \:. \label{QEDoffd}
\ee

When we compare these results with Eq.\ \eqref{Yres} for Yukawa theory,
the differences are the opposite sign of the ``spin-orbit'' term
[Eq.\ \eqref{Yspin} vs.\ Eq.\ \eqref{QEDspin}], the absence of retardation,
and, of course, the terms \eqref{QEDhf1}--\eqref{QEDhf3} which
introduce the hyperfine structure. The difference in signs of the
spin-orbit terms changes the qualitative features of the spectrum
drastically: nearly all energy corrections are negative, only
$\Delta E (1 \supfi{3} S_1)$ becomes positive for mass ratios $m_A/m_B$
close to one. The level ordering of the $2 P$ states is just opposite
to the Yukawa case. In the one-body limit, the characteristic
degeneracies appear, but in addition the $2 S$ states become degenerate 
with the $2 P$ states with $j_A = 1/2$ (compare with the discussion of the
one-body limit in the previous subsection). This is, of course, just the
famous degeneracy in orbital angular momentum characteristic of the
Coulomb potential (note that this degeneracy is broken in Yukawa theory by 
the retardation terms). Also, the $2 P$ states with $j_A = 1/2$ have
a lower energy in QED than the $j_A = 3/2$ states.

The hyperfine structure terms split the energies of the $1 S$ and the
$2 S$ states. Close to the one-body limit, we can again expand the
expressions \eqref{QEDres} in powers of $m_A/m_B$ to obtain the hyperfine
splittings (after diagonalizing the $2 \times 2$ matrix for the states
$2 \supfi{3} P_1$ and $2 \supfi{1} P_1$), which now appear to the order
$m_A/m_B$ (and $m_r \alpha^4$). Our results \eqref{QEDres} coincide
completely with the ones in Ref.\ \cite{BS77} close to the one-body limit
and for the case of equal masses \cite{Con03} (in the latter case, 
they coincide with the positronium results omitting the contributions from 
virtual annihilation there).
In fact, one can show that the expanded effective Schr\"odinger equation
in Eqs.\ \eqref{QEDkin}--\eqref{QEDhf3} is identical to the Pauli 
approximation of the Breit equation (and in the one-body limit to the Pauli 
approximation of the Dirac equation) \cite{BS77}. Depending on the form
in which the latter equations are written, the identification of terms may
require some algebraic labor. In particular, the Pauli approximation of
the Breit and Dirac equations are sometimes given in a not manifestly
hermitian form. In any case, the identity of the equations implies the
identity of the lowest-order relativistic corrections, for any state and
any mass ratio. We conclude that our approach reproduces the 
lowest-order fine and hyperfine structure
completely (at least) in the case of QED, from the effective Hamiltonian
($H_B$ or $H_W$) to order $H_1^2$, and it does so in a very transparent
and economic way.

To sum up, we have argued in this paper that we can obtain the lowest-order
fine and hyperfine structure in nearly nonrelativistic bound systems in a 
very straightforward way from the application of the generalized 
Gell-Mann--Low theorem. We have verified this claim for bound states in
QED, and we have presented, for the first time to our knowledge, results
for the fine and hyperfine structure of bound states in the Wick-Cutkosky
model and Yukawa theory. Although the results are physically appealing,
we have as yet no proof that our calculation of the fine and hyperfine
structure in the latter theories is complete. We have emphasized the
contributions from the retardation in the effective potential for the
Wick-Cutkosky model and Yukawa theory, and discussed the role of a peculiar 
antihermitian term that appears in this context, in particular with
respect to two possible choices $H_B$ and $H_W$ for the effective
Hamiltonian.

\subsection*{Acknowledgments}

The author is grateful to Norbert E. Ligterink for many interesting 
discussions on the subject and for help with the graphics. Financial 
support by CIC-UMSNH and Conacyt project 32729-E is acknowledged.

\begin{appendix}

\section{\label{formul} A collection of useful formulae}

We will present in this appendix the formulae that are used in the
actual analytic calculations of the perturbative corrections to the
Coulomb energy eigenvalues. Since we have done all of the calculations
directly in momentum space (as a warm-up for more complex calculations
in the future), we need, first of all, the Coulomb wave functions 
$\phi_{nlm} (\bf{p})$ in momentum space, here for the principal quantum 
numbers $n = 1, 2$:
\bal
\phi_{100} (\bf{p}) &= \frac{16 \pi \, (m_r \alpha)^{5/2}}
{\left[ (m_r \alpha)^2 + p^2 \right]^2} \, Y_{00} (\hat{\bf{p}}) 
\n \\
\phi_{200} (\bf{p}) &= \frac{\ds 4 \pi \sqrt{2} \, (m_r \alpha)^{5/2}
\left( p^2 - \frac{(m_r \alpha)^2}{4} \right)}{\ds \left( 
\frac{(m_r \alpha)^2}{4} + p^2 \right)^3} \, Y_{00} (\hat{\bf{p}}) \n \\
\phi_{21m} (\bf{p}) &= \frac{\ds 4 \pi \sqrt{\frac{2}{3}} \, 
(m_r \alpha)^{7/2} p}{\ds \left( \frac{(m_r \alpha)^2}{4} + p^2 \right)^3} \, 
Y_{1m} (\hat{\bf{p}}) \:. \label{phi0mom}
\eal
The wave functions are normalized to
\be
\int \frac{d^3 p}{(2 \pi)^3} \, \abs{\phi_{nlm} (\bf{p})}^2 = 1 \:.
\ee
In the cases of Yukawa theory and QED, the expressions above still have to 
be multiplied with the appropriate Pauli spinors for the spin orientation
of the fermions. It is convenient in the latter cases to use total angular
momentum eigenstates. In order to make contact to the usual spectroscopy,
we first couple the two spins 1/2 to a total spin $S = 0, 1$, and then
couple the total spin with the relative orbital angular momentum $l$
to the total angular momentum $J, M$. The explicit results for the
total angular momentum eigenstates 
$\supfi{2S + 1} \cal{Y}_{l M}^J (\hat{\bf{p}})$ are \cite{WL05}, in terms
of the well-known eigenstates $\chi_{S, m_S}$ of total spin:
\bal
\supfi{1} \cal{Y}_{J M}^J (\hat{\bf{p}}) &= Y_{J M} (\hat{\bf{p}}) \, 
\chi_{00} \:, \n \\[1mm]
\supfi{3} \cal{Y}_{J-1, M}^J (\hat{\bf{p}}) &= \frac{1}{\sqrt{2 J
(2J - 1)}} \left[ \sqrt{(J - M - 1)(J - M)} \, Y_{J - 1, M + 1} 
(\hat{\bf{p}}) \, \chi_{1, -1} \right. \n \\
&\phantom{=} 
+ \sqrt{2 (J - M)(J + M)} \, Y_{J - 1, M} (\hat{\bf{p}}) 
\, \chi_{10} 
\n \\ &\phantom{=} \left.
+ \sqrt{(J + M - 1)(J + M)} \, Y_{J - 1, M - 1}
(\hat{\bf{p}}) \, \chi_{11} \right]  \quad (J \ge 1) \:, \n \\[1mm]
\supfi{3} \cal{Y}_{J M}^J (\hat{\bf{p}}) &= \frac{1}{\sqrt{2 J (J + 1)}}
\left[ \sqrt{(J - M)(J + M + 1)} \, Y_{J, M + 1} (\hat{\bf{p}}) \, 
\chi_{1, -1} \right. \n \\
&\phantom{=} \left. + \sqrt{2} \, M \, Y_{J M} (\hat{\bf{p}}) \, \chi_{10}
- \sqrt{(J - M + 1)(J + M)} \, Y_{J, M - 1} 
(\hat{\bf{p}}) \, \chi_{11} \right] \quad (J \ge 1) \:, \n \\[1mm]
\supfi{3} \cal{Y}_{J + 1, M}^J (\hat{\bf{p}}) &=
\frac{1}{\sqrt{2(J + 1) (2 J + 3)}}
\left[ \sqrt{(J + M + 1)(J + M + 2)} \, Y_{J + 1, M + 1} 
(\hat{\bf{p}}) \, \chi_{1, -1} \right. \n \\
&\phantom{=} 
- \sqrt{2 (J - M + 1)(J + M + 1)} \, Y_{J + 1, M} 
(\hat{\bf p}) \, \chi_{10} 
\n \\ &\phantom{=} \left.
+ \sqrt{(J - M + 1)(J - M + 2)} \, Y_{J + 1, M - 1} 
(\hat{\bf{p}}) \, \chi_{11} \right] \:.
\eal
For the calculation of the perturbative corrections in these theories,
one needs to apply the operators $(\bf{p} \cdot \bm{\sigma}_{A,B})$ to these
total angular momentum eigenstates. The corresponding formulae were also
worked out in Ref.\ \cite{WL05}. They read
\bal
(\bf{p} \cdot \bm{\sigma}_A) \, \supfi{1} \cal{Y}_{J M}^J
(\hat{\bf{p}}) &= p \left( \sqrt{\frac{J}{2 J + 1}} \, 
\supfi{3}\cal{Y}_{J - 1, M}^J (\hat{\bf{p}}) - \sqrt{\frac{J + 1}{2 J + 1}} 
\, \supfi{3} \cal{Y}_{J + 1, M}^J (\hat{\bf{p}}) \right) \:, \n \\
(\bf{p} \cdot \bm{\sigma}_A) \, \supfi{3} \cal{Y}_{J - 1, M}^J
(\hat{\bf{p}}) &= p \left( \sqrt{\frac{J}{2 J + 1}} \, 
\supfi{1} \cal{Y}_{J M}^J (\hat{\bf{p}}) - \sqrt{\frac{J + 1}{2 J + 1}} \, 
\supfi{3} \cal{Y}_{J M}^J (\hat{\bf{p}}) \right) \:, \n \\
(\bf{p} \cdot \bm{\sigma}_A) \, \supfi{3} \cal{Y}_{J M}^J
(\hat{\bf{p}}) &= - p \left( \sqrt{\frac{J + 1}{2 J + 1}} \, 
\supfi{3} \cal{Y}_{J - 1, M}^J (\hat{\bf{p}}) + \sqrt{\frac{J}{2 J + 1}} \, 
\supfi{3} \cal{Y}_{J + 1, M}^J (\hat{\bf{p}}) \right) \:, \n \\
(\bf{p} \cdot \bm{\sigma}_A) \, \supfi{3} \cal{Y}_{J + 1, M}^J
(\hat{\bf{p}}) &= - p \left( \sqrt{\frac{J + 1}{2 J + 1}} \, 
\supfi{1} \cal{Y}_{J M}^J (\hat{\bf{p}}) + \sqrt{\frac{J}{2 J + 1}} \, 
\supfi{3} \cal{Y}_{J M}^J (\hat{\bf{p}}) \right) \:, 
\eal 
and
\bal
(\bf{p} \cdot \bm{\sigma}_B) \, \supfi{1} \cal{Y}_{J M}^J
(\hat{\bf{p}}) &= - p \left( \sqrt{\frac{J}{2 J + 1}} \, 
\supfi{3} \cal{Y}_{J - 1, M}^J (\hat{\bf{p}}) - \sqrt{\frac{J + 1}{2 J + 1}} 
\, \supfi{3} \cal{Y}_{J + 1, M}^J (\hat{\bf{p}}) \right) \:, \n \\
(\bf{p} \cdot \bm{\sigma}_B) \, \supfi{3} \cal{Y}_{J - 1, M}^J
(\hat{\bf{p}}) &= - p \left( \sqrt{\frac{J}{2 J + 1}} \, 
\supfi{1} \cal{Y}_{J M}^J (\hat{\bf{p}}) + \sqrt{\frac{J + 1}{2 J + 1}} \, 
\supfi{3} \cal{Y}_{J M}^J (\hat{\bf{p}}) \right) \:, \n \\
(\bf{p} \cdot \bm{\sigma}_B) \, \supfi{3} \cal{Y}_{J M}^J
(\hat{\bf{p}}) &= - p \left( \sqrt{\frac{J + 1}{2 J + 1}} \, 
\supfi{3} \cal{Y}_{J - 1, M}^J (\hat{\bf{p}}) + \sqrt{\frac{J}{2 J + 1}} \, 
\supfi{3} \cal{Y}_{J + 1, M}^J (\hat{\bf{p}}) \right) \:, \n \\
(\bf{p} \cdot \bm{\sigma}_B) \, \supfi{3} \cal{Y}_{J + 1, M}^J
(\hat{\bf{p}}) &= p \left( \sqrt{\frac{J + 1}{2 J + 1}} \, 
\supfi{1} \cal{Y}_{J M}^J (\hat{\bf{p}}) - \sqrt{\frac{J}{2 J + 1}} \, 
\supfi{3} \cal{Y}_{J M}^J (\hat{\bf{p}}) \right) \:.
\eal 
In the special case $J=0$, the states $\supfi{3} \cal{Y}_{J - 1, M}^J 
(\hat{\bf p})$ and $\supfi{3} \cal{Y}_{J M}^J (\hat{\bf p})$ do not exist, 
and on the right-hand sides for the application of one of the helicity 
operators to $\supfi{1} \cal{Y}_{J M}^J (\hat{\bf p})$ and
$\supfi{3} \cal{Y}_{J + 1, M}^J (\hat{\bf p})$, only one term remains.
For the application of $(\bm{\sigma}_A \cdot \bm{\sigma}_B)$, we use
the well-known identity
\be
(\bm{\sigma}_A \cdot \bm{\sigma}_B) \supfi{2S + 1} \cal{Y}_{l M}^J
(\hat{\bf{p}}) = \left[ 2 S (S + 1) - 3 \right] 
\supfi{2S + 1} \cal{Y}_{l M}^J (\hat{\bf{p}}) \:.
\ee

After the application of these formulae, the angular integrations can
be performed with the help of the partial wave decomposition and the
spherical harmonics addition theorem, Eq.\ \eqref{addtheorem}. To this
end, the coefficient functions have to be calculated from Eq.\ 
\eqref{pwcoeff} which is elementary in all relevant cases. 

Finally, the integrations over $p$ and $p'$ have to be performed. All
appearing integrals are again elementary, the only exception being
integrals of the type
\be
\int_0^\infty d p \, \frac{p^{2k+1}}{(\lambda^2 + p^2)^{n+1}} 
\ln \left( \frac{p + p'}{|p - p'|} \right) \:,
\ee
with $n=1,2,3,\ldots$ and $0 \le k \le n$. All these integrals
can be obtained by differentiation with respect to $\lambda^2$ and
trivial algebraic manipulations of the fraction from 
\be
I = \int_0^\infty dp \, \frac{p}{\lambda^2 + p^2} \ln \left(
\frac{p + p'}{|p - p'|} \right) \:,
\ee
which we will now evaluate. To this end, consider the analytic function
\be
F(p) = \frac{p}{\lambda^2 + p^2} \ln \left( \frac{p + p'}{p - p'} \right)
\ee
which is even under $p \to -p$ and its real part coincides with the
integrand of $I$, so that
\be
I = \frac{1}{2} \, \text{Re} \int_{-\infty}^\infty d p \, F(p) \:.
\ee
As far as its analytic structure is concerned, $F(p)$ has simple poles at
$p = \pm i \lambda$ and a cut on the real axis from $-p'$ to $p'$. We can
hence evaluate $I$ by integrating along the real axis and (e.g.) the upper
rim of the cut and closing the contour through the upper infinite semicircle.
Then only the pole at $i \lambda$ is picked up (suppose $\lambda > 0$), and
the result is
\be
I = \pi \arctan (p'/\lambda) \:.
\ee

\section{\label{FWtrans} A transformation of Foldy-Wouthuysen type}

In this appendix we will devise an alternative method for the calculation
of the contributions to the energy corrections that arise from the
antihermitian term \eqref{WCreta} or \eqref{Yreta} in the Wick-Cutkosky model 
or Yukawa theory, respectively. Unlike the ``direct'' calculation in
Section \ref{PTat}, we will construct a similarity transformation that
converts $H_B$ to a hermitian operator, to the perturbative order
required. Since the eigenvalues are unchanged under the similarity
transformation, we can then use common ``hermitian'' perturbation theory 
to calculate the corrections to the energy. 

In fact, in Section \ref{OHder} a similarity transformation was
introduced that does just this, see Eq.\ \eqref{defHW}. Although we
will take the form of this transformation as a motivation, the
transformation \eqref{defHW} as it stands is plagued by infrared 
divergencies. For this reason we rather choose a transformation of
the Foldy-Wouthuysen type because in this case the 
Baker-Campell-Hausdorff formula guarantees a convenient commutator
structure for the transformed operator. Since the similarity transform
has to map a nonhermitian operator to a hermitian one (at least to a certain
perturbative order), it can obviously not be unitary. We hence consider
transformations of the general form
\be
S = \exp \left( - T^{\ssc (1)} - T^{\ssc (2)} - \ldots \right) : \Omega_0 
\to \Omega_0 \:, \label{SFW}
\ee
where the operator $T^{\ssc (2)}$ is of higher perturbative order than
$T^{\ssc (1)}$, etc. It follows for the similarity transformed Hamiltonian
\be
H_B' = S^{-1} H_B S = H_B + \left[ T^{\ssc (1)}, H_B \right] + \frac{1}{2}
\big[ T^{\ssc (1)}, \left[ T^{\ssc (1)}, H_B \right] \big] +
\left[ T^{\ssc (2)}, H_B \right] + \ldots \:.  \label{HBFW}
\ee

We could now take the similarity transformation to coincide with the one in
Eq.\ \eqref{defHW} to lowest nontrivial order in $g$,
\be
S' = ( U_B^\dagger U_B )^{-1/2} + \cal{O} (g^4) \:, \label{Sans}
\ee
by choosing
\be
T^{\prime \ssc (1)} = \frac{1}{2} \int_{-\infty}^0 d t_1 \int_{-\infty}^0 
d t_2 \, e^{-\epsilon (\abs{t_1} +\abs{t_2})} P_0 H_1 (t_1) H_1 (t_2) P_0 
\ee
[cf.\ Eq.\ \eqref{UB2}]. However, this is still a quite complicated choice 
because of the infrared divergencies mentioned before, so we will
take a somewhat simpler alternative and leave the diagonal
contributions (vacuum and kinetic energies) out of $T^{\prime \ssc (1)}$, 
defining $T^{\ssc (1)}$ explicitly through its matrix elements in the 
center-of-mass system as
\bal
\lefteqn{\langle \bf{p}, r, s | T^{\ssc (1)} | \bf{p}', r', s' \rangle} 
\n \\
&= - \frac{g^2}{2 \sqrt{2 E_{\bf{p}}^A \, 2 E_{\bf{p}}^B \, 
2 E_{\bf{p}'}^A \, 2 E_{\bf{p}'}^B}} \,
\frac{\left[ \bar{u}_A (\bf{p}, r) \, u_A (\bf{p}', r') \right] 
\left[ \bar{u}_B (- \bf{p}, s) \, u_B (- \bf{p}', s') \right]}
{E_{\bf{p}}^A + E_{\bf{p}}^B - E_{\bf{p}'}^A - E_{\bf{p}'}^B} \n \\
&\phantom{=} \times
\frac{1}{2 \sqrt{(\bf{p} - \bf{p}')^2 + \mu^2}} \left( \frac{1}{E_{\bf{p}}^A 
+ \sqrt{(\bf{p} - \bf{p}')^2 + \mu^2} - E_{\bf{p}'}^A} + 
\frac{1}{E_{\bf{p}}^B + \sqrt{(\bf{p} - \bf{p}')^2 + \mu^2} - E_{\bf{p}'}^B} 
\right.\n \\[2mm]
&\phantom{=} \hspace{1cm} \left. - \frac{1}{E_{\bf{p}'}^A +
\sqrt{(\bf{p} - \bf{p}')^2 + \mu^2} - E_{\bf{p}}^A} - 
\frac{1}{E_{\bf{p}'}^B + \sqrt{(\bf{p} - \bf{p}')^2 + \mu^2} - E_{\bf{p}}^B}
\right) \label{T1def}
\eal
Definition \eqref{T1def} properly applies to Yukawa theory, for the 
Wick-Cutkosky model one just has to omit the products of Dirac spinors. 
We have introduced a mass $\mu$ for the exchanged boson as an IR regulator 
(see below).

Equation \eqref{HBFW} above, to lowest nontrivial order, then leads to
\be
H_B' = H_B + \left[ T^{\ssc (1)}, H_0 \right] + \cal{O} (g^4)
= H_W + \cal{O} (g^4) \:, \label{HBtrans}
\ee
just as for the full transformation in Eq.\ \eqref{HWcom}. 
However, while we were content in Section \ref{OHder} with calculating $H_W$
strictly to order $H_1^2$, for the lowest-order fine
and hyperfine structure we need all terms of order $m_r \alpha^4$, and
such terms may arise from the order-$g^4$ contibutions to $H_B'$. In
fact, we will now show that this is the case, which means that
$H_B' \neq H_W$ to the order $m_r \alpha^4$.

In order to extract the terms of order $m_r \alpha^4$ from the order-$g^4$
terms in Eq.\ \eqref{HBFW}, it is sufficient to work with $T^{\ssc (1)}$
to lowest order in a momentum expansion, explicitly
\be
\langle \bf{p}, r, s | T^{\ssc (1)} | \bf{p}', r', s' \rangle
= \frac{2 \pi \alpha}{\left[ (\bf{p} - \bf{p}')^2 + \mu^2 \right]^{3/2}}
\, \delta_{r r'} \delta_{s s'} + \cal{O} (\alpha^3) \label{T1exp}
\ee
(for Yukawa theory, for the Wick-Cutkosky model just omit the
Kronecker deltas). For the discussion of IR convergence or divergence, 
as well as for the estimation of the order in $\alpha$ by IR power counting, 
we have to take into account that Eq.\ \eqref{T1exp} is integrated
over $d^3 p'$ when it is applied to a function of $\bf{p}'$. This adds three
powers of $\alpha$ to the power counting, so that the leading term in
Eq.\ \eqref{T1exp} is IR divergent and of order $\alpha$ (the IR regulator 
is ignored in the power counting). The fact that $T^{\ssc (1)}$ is IR 
divergent to lowest order is uncomfortable. Incidentally, given the
relation of $T^{\ssc (1)}$ to $( U_B^\dagger U_B )^{-1/2}$ and hence
to $U_W$, this IR divergence sheds doubts on the proper existence of
the Okubo map. We have not completely analyzed the question of the
IR convergence or divergence of $U_W$ yet. In any event, the transformed 
Hamiltonian will turn out to be IR convergent to the order $m_r \alpha^4$
considered at present.

Now consider the momentum expansion of
\bal
\lefteqn{\langle \bf{p}, r, s | H_B + \frac{1}{2} \left[ T^{\ssc (1)}, H_0 
\right] | \bf{p}', r', s' \rangle} \hspace{5mm} \n \\ 
&= \langle \bf{p}, r, s | H_0 | \bf{p}', r', s' \rangle 
- \frac{4 \pi \alpha}{(\bf{p} - \bf{p}')^2 + \mu^2} \left( 1
- \frac{1}{8 m_r} \frac{p^2 - p^{\prime 2}}
{\sqrt{(\bf{p} - \bf{p}')^2 + \mu^2}} \right) \delta_{r r'} \delta_{s s'} + 
\cal{O} (m_r \alpha^4) \label{HBcexp}
\eal
(again omitting the Kronecker deltas for the case of the Wick-Cutkosky
model). Observe that all the terms are IR convergent by power counting,
hence the IR cutoff $\mu$ is not stictly necessary here.
When the commutator with $T^{\ssc (1)}$ is taken in Eq.\ \eqref{HBFW}, the 
term with $H_0$ on the right-hand-side of Eq.\ \eqref{HBcexp}
contributes to $H_W$ in Eq.\ \eqref{HBtrans}, the following term of order 
$m_r \alpha^2$ does not contribute at all up to orders $m_r \alpha^4$, and 
it is the last term of order $m_r \alpha^3$ which gives the important 
order-$m_r \alpha^4$ contribution with matrix elements
\be
- \frac{2 \pi^2 \alpha^2}{m_r} \int \frac{d^3 q}{(2 \pi)^3}
\frac{(\bf{p} - \bf{q}) \cdot (\bf{q} - \bf{p}')}{\left[ 
(\bf{p} - \bf{q})^2 + \mu^2 \right]^{3/2} \left[ 
(\bf{q} - \bf{p}')^2 + \mu^2 \right]^{3/2}} \, \delta_{r r'} \delta_{s s'} \:.
\ee
The latter expression is IR finite by power counting, hence the limit
$\mu \to 0$ can safely be taken. It is then simple to perform the
$d^3 q$-integration, with the final result
\be
\langle \bf{p}, r, s | H_B' | \bf{p}', r', s' \rangle =
\langle \bf{p}, r, s | H_W | \bf{p}', r', s' \rangle + \frac{\alpha^2}{m_r} 
\frac{1}{\abs{\bf{p} - \bf{p}'}} \, \delta_{r r'} \delta_{s s'}
+ \cal{O} (m_r \alpha^5) \label{HBprime}
\ee
(without the Kronecker deltas for the Wick-Cutkosky model, as usual).
This result coincides with the calculation in second-order ``antihermitian''
perturbation theory, see Eq.\ \eqref{Eantiopm}. The possible physical
significance of the ``new'' $m_r \alpha^4$-term is discussed at the end
of Subsection \ref{PTat}. It appears plausible from the result \eqref{HBprime}
and the consistency of the general formalism that this additional term is
also generated by the similarity transformation $( U_B^\dagger U_B )^{-1/2}$.

To close this appendix, let us have a look at the contributions to order
$m_r \alpha^5$ in $H_B'$. It is out of the question to present any
details of the corresponding lengthy calculations, hence we will be
content here with quoting the main results. First of all, it can be
shown that all contributions of this order come from the next-higher terms
in the expansions \eqref{T1exp} and \eqref{HBcexp}, inserted in the
commutator
\be
\left[ T^{\ssc (1)}, H_B + \frac{1}{2} \left[ T^{\ssc (1)}, H_0 
\right] \right] \:.
\ee
The complete result is cumbersome, but we will concentrate here on the
IR divergent contributions that potentially
invalidate the similarity transformation. They are explicitly given by
the antihermitian expression
\be
\langle \bf{p}, r, s | H_B' | \bf{p}', r', s' \rangle \sim
\frac{4 \pi^2 \alpha^2}{3 m_r^2} \int \frac{d^3 q}{(2 \pi)^3}
\frac{p^2 - p^{\prime 2}}{\left[ (\bf{p} - \bf{q})^2 + \mu^2 \right]^{3/2} 
\left[ (\bf{q} - \bf{p}')^2 + \mu^2 \right]} \, \delta_{r r'} 
\delta_{s s'} + \cal{O} (m_r \alpha^6) \:,
\ee
where ``$\sim$'' refers to the equality of the IR divergent terms.
It is now possible to remove these terms from $H_B'$ by introducing a 
further operator $T^{\ssc (2)}$ in the similarity transformation \eqref{SFW}, 
defined by
\be
\langle \bf{p}, r, s | T^{\ssc (2)} | \bf{p}', r', s' \rangle
= \frac{8 \pi^2 \alpha^2}{3 m_r} \int \frac{d^3 q}{(2 \pi)^3}
\frac{1}{\left[ (\bf{p} - \bf{q})^2 + \mu^2 \right]^{3/2} 
\left[ (\bf{q} - \bf{p}')^2 + \mu^2 \right]} \, \delta_{r r'} 
\delta_{s s'} \:.
\ee
It is conceivable that all possible IR divergencies can be removed in 
this way, at least recursively order by order in $\alpha$. Note, however,
that $T^{\ssc (1)}$ as well as $T^{\ssc (2)}$ are IR divergent themselves.

\end{appendix}


\begin{thebibliography}{99}
\bibitem{Web00} A. Weber, in \textit{Particles and Fields --- Seventh Mexican 
Workshop}, edited by A. Ayala, G. Contreras, and G. Herrera, AIP Conf.
Proc. No. 531 (AIP, New York, 2000), p. 305, hep-th/9911198.
\bibitem{WL02} A. Weber and N.E. Ligterink, Phys. Rev. D \textbf{65},
025009 (2002).
\bibitem{WL05} A. Weber and N.E. Ligterink, hep-ph/0506123.
\bibitem {Web03} A. Weber, in \textit{Quark Confinement and the Hadron
Spectrum V}, edited by N. Brambilla and G.M. Prosperi (World Scientific,
Singapore, 2003), p. 428.
\bibitem{BS77} H.A. Bethe and E.E. Salpeter, \textit{Quantum  mechanics of 
one- and two-electron atoms} (Plenum Press, New York, 1977).
\bibitem{Bre29} G. Breit, Phys. Rev. \textbf{34}, 553 (1929).
\bibitem{BS51} E.E. Salpeter and H.A. Bethe, Phys. Rev. \textbf{84},
1232 (1951).
\bibitem{GL51} M. Gell-Mann and F. Low, Phys. Rev. \textbf{84},
350 (1951).
\bibitem{Sal52} E.E. Salpeter, Phys. Rev. \textbf{87}, 328 (1952).
\bibitem{WC54} G.C. Wick, Phys. Rev. \textbf{96}, 1124 (1954);
R.E. Cutkosky, \textit{ibid.} \textbf{96}, 1135 (1954).
\bibitem{FH73} G. Feldman, T. Fulton, and J. Townsend, Phys. Rev. D
\textbf{7}, 1814 (1973).
\bibitem{Blo58} C. Bloch, Nucl. Phys. \textbf{6}, 329 (1958).
\bibitem{Wil70} K.G. Wilson, Phys. Rev. D \textbf{2}, 1438 (1970).
\bibitem{GH76} M. Gari and H. Hyuga, Z. Phys. A \textbf{277}, 291 (1976);
M. Gari, G. Niephaus, and B. Sommer, Phys. Rev. C \textbf{23}, 504 (1981)
\bibitem{KG99} A. Kr\"uger and W. Gl\"ockle, Phys. Rev. C
\textbf{59}, 1919 (1999).
\bibitem{Oku54} S. Okubo, Prog. Theor. Phys. \textbf{12}, 603 (1954).
\bibitem{DL55} A. Dalgarno and J.T. Lewis, Proc. R. Soc. (London), Ser.
A \textbf{233}, 70 (1955).
\bibitem{GR00} I.S. Gradshteyn and I.M. Ryzhik, \textit{Table of Integrals, 
Series, and Products} (Academic Press, San Diego, 2000), sixth edition.
\bibitem{Gro74} F. Gross, Phys. Rev. D \textbf{10}, 223 (1974).
\bibitem{GVH92} F. Gross, J.W. Van Orden, and K. Holinde, Phys. Rev. C 
\textbf{45}, 2094 (1992).
\bibitem{LB01} N.E. Ligterink and B.L.G. Bakker, hep-ph/0010167. 
\bibitem{WLS00} A. Weber, J.C. L\'opez Vieyra, C.R. Stephens, S. Dilcher, 
and P.O. Hess, Int. J. Mod. Phys. \textbf{A16}, 4377 (2001).
\bibitem{MC00} M. Mangin-Brinet and J. Carbonell, Phys. Lett. B 
\textbf{474}, 237 (2000).
\bibitem{GRS95} O.W. Greenberg, R. Ray, and F. Schlumpf, Phys. Lett. B
\textbf{353}, 284 (1995).
\bibitem{Con03} Y. Concha S\'anchez, B.Sc. thesis, Universidad Michoacana
de San Nicol\'as de Hidalgo, 2003.
\end{thebibliography}
\end{document}